# Cluster decay half lives of trans-lead nuclei within the Coulomb and proximity potential model

K. P. Santhosh*, B. Priyanka and M. S. Unnikrishnan

*School of Pure and Applied Physics, Kannur University, Payyanur Campus, Payyanur 670 327, India*

**Abstract**

Within the Coulomb and proximity potential model (CPPM) the cluster decay process in $^{199-226}$Fr, $^{206-232}$Ac, $^{209-237}$Th, $^{212-238}$Pa, $^{217-241}$U, $^{225-242}$Np, $^{225-244}$Pu, $^{231-246}$Am, $^{202-230}$Ra and $^{233-249}$Cm isotopes leading to the doubly magic $^{208}$Pb and neighbouring nuclei are studied. The computed half lives are compared with available experimental data and are in good agreement with each other. The half lives are also computed using the Universal formula for cluster decay (UNIV) of Poenaru et al, Universal decay law (UDL) and the scaling law of Horoi et al, and their comparisons with CPPM values are found to be in agreement. The calculations for the emission of $^{22}$O, $^{20}$O, $^{20}$O from the parents $^{209-237}$Th, $^{202-230}$Ra and $^{217-240}$U respectively were the experimental values are not available are also done. It is found that most of the decay modes are favourable for measurement $(T_{1/2} < 10^{30} s)$, and this observation will serve as a guide to the future experiments. The odd-even staggering (OES) are found to be more prominent in the emission of odd mass clusters. The Geiger – Nuttall plots of $\log_{10}(T_{1/2})$ vs. $Q^{-1/2}$ for various clusters ranging from $^{14}$C to $^{34}$Si from different isotopes of heavy parent nuclei with atomic numbers within the range $87 \leq Z \leq 96$ have been studied and are found to be linear. Our study reveals the role of doubly magic $^{208}$Pb daughter nuclei in cluster decay process and also reveal the fact that the role of neutron shell closure is crucial than proton shell closure.

*Tel: +919495409757; Fax: +914972806402
*Email address*: drkpsanthosh@gmail.com

## 1. Introduction

Cluster radioactivity, the rare, cold process intermediate between alpha decay and spontaneous fission is now a well established phenomenon both from the experimental and the theoretical side. The rare nature of this process is due to the fact that cluster radioactivity is masked by several alpha emissions. Exotic decay or cluster radioactivity was first predicted by Sandulescu et al., [1] in 1980 and such decays was first observed experimentally by Rose and Jones [2] in 1984 in the radioactive decay of $^{223}$Ra by the emission of $^{14}$C with a half life of 3.7 ± 1.1 years. Later on, several clusters were observed experimentally from various parents in the trans-lead region with partial half lives from $10^{11}$ up to $10^{30}$ s and branching ratios relative to alpha decay from $10^{-9}$ down to $10^{-19}$. At present about 20 clusters ranging from $^{14}$C to $^{34}$Si has been confirmed to be emitted from parent nuclei ranging from $^{221}$Fr to $^{242}$Cm [3].

As mentioned earlier, cluster decay is well known to be a process intermediate between alpha particle decay and spontaneous fission. These two extreme processes of hadronic decay of nuclei are described using completely different formalisms. The first one ("alpha decay"-like) is considered to be non-adiabatic [4, 5] and the second one ("fission"-like), on the contrary is considered to be adiabatic [6, 7]. The phenomenon of cluster radioactivity has been explained making use of several theoretical models, both alpha like approach [8] and fission like approach [9, 10, 11]. In the former, the probability of cluster formation is determined by the overlap of the parent nucleus wave function with those of both fragments resulting in a sudden formation of a cluster which then attempts to penetrate the Coulomb barrier. In the later, cluster decay is considered to be a single step process. It includes the pre-scission phase where the fragments are overlapping. Here, an exotic nucleus is considered to split up into two asymmetric fragments. It was the pioneering works by Poenaru [12] which revealed that the transmission probability through that part of the potential barrier before the saddle point, i.e. the pre-scission phase, simulates the spectroscopic factor present in the "α decay"-like models, which is implicitly assumed to be unity in the "fission"-like models.

The emissions of heavier clusters, such as $^{14}$C, $^{20}$O, $^{24}$Ne, $^{28}$Mg and $^{32}$Si, have been established experimentally in trans-lead nuclei decaying into daughters around the doubly magic nucleus of $^{208}$Pb [2, 13, 14]. Later a second island of heavy-cluster radio activities was predicted [15] in trans-tin nuclei decaying into daughters close to the doubly magic nucleus of $^{100}$Sn. The spontaneous emission of a heavier cluster, namely $^{24}$Ne from $^{231}$Pa, $^{233}$U and $^{230}$Th, was detected by Sandulescu et al [16, 17] in Dubna. It was Barwick et al [18] who first experimentally confirmed the emission of $^{24}$Ne cluster from $^{232}$U. Bonetti et al [19] reported neon radioactivity of uranium isotopes ($^{232,234,235}$U) and Price et al [20] experimentally observed the emission of $^{23}$F and $^{24}$Ne emission from $^{231}$Pa nucleus. Bonetti et al [21] have experimentally studied the cluster decay of $^{230}$U isotope with the emission of neon clusters ($^{22}$Ne and $^{24}$Ne). The $^{14}$C decay of $^{221}$Fr, $^{221-224,226}$Ra and $^{225}$Ac; $^{20}$O decay of $^{228}$Th; $^{23}$F decay of $^{231}$Pa; $^{24}$Ne decay of $^{230}$Th, $^{231}$Pa and $^{232-234}$U; $^{28}$Mg decay of $^{234}$U and $^{236,238}$Pu; $^{30}$Mg decay of $^{238}$Pu; $^{32}$Si decay of $^{238}$Pu, and $^{34}$Si radioactivity of $^{242}$Cm have been identified experimentally.

By considering the interacting potential as the sum of Coulomb and Proximity potential (PPCPM) Shi and Swiatecki [10, 22] have studied exotic decay of some heavy nuclei in trans-lead region. Using Analytical super asymmetric fission model (ASAFM) Poenaru et al [9, 23, 24] have calculated half-lives for several cluster decay modes of some heavy nuclei in the trans-lead region. Since the first experimental observation of cluster radioactivity in 1984, the ASAFM have been successfully used to compute half life for alpha and cluster radioactivity in heavy and superheavy nuclides (see the reviews [25, 26] and references therein). Recently Poenaru et al [27] in their letter predicted heavy particle radioactivity (HPR) from elements with Z > 110 leading to doubly magic $^{208}$Pb with shorter half life and larger branching ratio relative to alpha decay. Shanmugam et al [28] have calculated half-lives for exotic decay of some experimentally observed decay modes by using cubic plus Yukawa plus exponential model (CYEM). Buck et al [29, 30] have calculated half-lives for some of the experimentally observed decay modes using unifying model for alpha and exotic decay. Using preformed cluster model (PCM), Malik et al [8] have studied the cluster decay of some experimentally observed decay modes. In subsequent years, the authors [31-35] have extensively studied the cluster decay of heavy nuclei in the trans-lead region.

Santhosh et al [36-38] have calculated half-lives for experimentally observed cluster decay modes of several heavy nuclei in the trans-lead and trans-tin region by using Coulomb Proximity potential model (CPPM). Tavares et al [39] have investigated exotic decay of heavy nuclei within the framework of a semi-empirical, one parameter model based on a quantum mechanical, tunnelling mechanism through a potential barrier by taking into account both centrifugal and overlapping effects in half-life evaluations. Warda and Robledo [40] have recently studied cluster radioactivity of actinide nuclei by using mean-field Hartree-Fock-Bogoliubov theory with the phenomenological Gogny interaction.

In the present paper we have investigated the cluster decay process in $^{199-226}$Fr, $^{206-232}$Ac, $^{209-237}$Th, $^{212-238}$Pa, $^{217-241}$U, $^{225-242}$Np, $^{225-244}$Pu, $^{231-246}$Am, $^{202-230}$Ra and $^{233-249}$Cm isotopes leading to doubly magic $^{208}$Pb and neighbouring nuclei. We have considered all the parent-cluster combinations, where the experimental results were available and we would confidently like to mention here that this is a unified theoretical study done in this manner. The comparisons of our calculations with the values obtained using Universal (UNIV) curve of Poenaru et al [41], Universal decay law (UDL) of Qi et al [42] and the scaling law of Horoi et al [43] is also done. We have also performed the calculations for the emission of

$^{22}$O, $^{20}$O, $^{20}$O from the parents $^{209-237}$Th, $^{202-230}$Ra and $^{217-240}$U respectively were the experimental values are not available.

The Coulomb and Proximity Potential Model (CPPM) is presented in detail in section 2, results and discussions on the cluster decay of the nuclei under study is given in section 3 and a conclusion on the entire work is given in section 4.

## 2. The Coulomb and proximity potential model (CPPM)

In the Coulomb and proximity potential model (CPPM), the potential energy barrier is taken as the sum of Coulomb potential, proximity potential and centrifugal potential for the touching configuration and for the separated fragments. For the pre-scission (overlap) region, simple power law interpolation as done by Shi and Swiatecki [10] is used. The inclusion of proximity potential reduces the height of the potential barrier, which closely agrees with the experimental result. The proximity potential was first used by Shi and Swiatecki [10] in an empirical manner and has been quite extensively used over a decade by Gupta *et al.*, [8] in the Preformed Cluster Model (PCM). R K Puri *et al.*, [44, 45] has been using different versions of proximity potential for studying fusion cross section of different target-projectile combinations. In our model contribution of both internal and external part of the barrier is considered for the penetrability calculation. In present model assault frequency, $\nu$ is calculated for each parent-cluster combination which is associated with vibration energy. But Shi and Swiatecki [46] get $\nu$ empirically, unrealistic values $10^{22}$ for even-A parents and $10^{20}$ for odd-A parents.

The interacting potential barrier for two spherical nuclei is given by

$$V = \frac{Z_1 Z_2 e^2}{r} + V_p(z) + \frac{\hbar^2 \ell(\ell+1)}{2\mu r^2} \qquad \text{, for } z > 0 \qquad (1)$$

Here $Z_1$ and $Z_2$ are the atomic numbers of the daughter and emitted cluster, 'z' is the distance between the near surfaces of the fragments, 'r' is the distance between fragment centres, $\ell$ represents the angular momentum, $\mu$ the reduced mass, $V_P$ is the proximity potential given by Blocki *et al.*, [47] as

$$V_p(z) = 4\pi\gamma b \left[ \frac{C_1 C_2}{(C_1 + C_2)} \right] \Phi\left(\frac{z}{b}\right) \quad (2)$$

With the nuclear surface tension coefficient,

$$\gamma = 0.9517 [1 - 1.7826(N-Z)^2 / A^2] \quad \text{MeV/fm}^2 \quad (3)$$

where N, Z and A represent the neutron, proton and mass number of the parent, $\Phi$ represents the universal proximity potential [48] given as

$$\Phi(\varepsilon) = -4.41 e^{-\varepsilon/0.7176}, \text{ for } \varepsilon \geq 1.9475 \quad (4)$$

$$\Phi(\varepsilon) = -1.7817 + 0.9270\varepsilon + 0.0169\varepsilon^2 - 0.05148\varepsilon^3, \text{ for } 0 \leq \varepsilon \leq 1.9475 \quad (5)$$

with $\varepsilon = z/b$, where the width (diffuseness) of the nuclear surface $b \approx 1$ and Süsmann central radii $C_i$ of the fragments related to sharp radii $R_i$ is

$$C_i = R_i - \left(\frac{b^2}{R_i}\right) \quad (6)$$

For $R_i$ we use the semi empirical formula in terms of mass number $A_i$ as [47]

$$R_i = 1.28 A_i^{1/3} - 0.76 + 0.8 A_i^{-1/3} \quad (7)$$

The potential for the internal part (overlap region) of the barrier is given as

$$V = a_0 (L - L_0)^n \quad \text{for } z < 0 \quad (8)$$

where $L = z + 2C_1 + 2C_2$ and $L_0 = 2C$, the diameter of the parent nuclei. The constants $a_0$ and n are determined by the smooth matching of the two potentials at the touching point.

Using one dimensional WKB approximation, the barrier penetrability P is given as

$$P = \exp\left\{ -\frac{2}{\hbar} \int_a^b \sqrt{2\mu(V-Q)} \, dz \right\} \quad (9)$$

Here the mass parameter is replaced by $\mu = mA_1A_2/A$, where m is the nucleon mass and $A_1$, $A_2$ are the mass numbers of daughter and emitted cluster respectively. The turning points "a" and "b" are determined from the equation, $V(a) = V(b) = Q$. The above integral can be evaluated numerically or analytically, and the half life time is given by

$$T_{1/2} = \left(\frac{\ln 2}{\lambda}\right) = \left(\frac{\ln 2}{\upsilon P}\right) \qquad (10)$$

where, $\upsilon = \left(\frac{\omega}{2\pi}\right) = \left(\frac{2E_v}{h}\right)$ represent the number of assaults on the barrier per second and $\lambda$ the decay constant. $E_v$, the empirical vibration energy is given as [49]

$$E_v = Q\left\{0.056 + 0.039\exp\left[\frac{(4 - A_2)}{2.5}\right]\right\} \qquad , \quad \text{for} \quad A_2 \geq 4 \qquad (11)$$

In the classical method, the $\alpha$ particle is assumed to move back and forth in the nucleus and the usual way of determining the assault frequency is through the expression given by $v = velocity/(2R)$, where R is the radius of the parent nuclei. But the alpha particle has wave properties; therefore a quantum mechanical treatment is more accurate. Thus, assuming that the alpha particle vibrates in a harmonic oscillator potential with a frequency $\omega$, which depends on the vibration energy $E_v$, we can identify this frequency as the assault frequency $v$ given in eqns. (10)-(11).

### 3. Results and discussions

The cluster decay half lives in the emission of clusters $^{14}C$, $^{15}N$, $^{18, 20, 22}O$, $^{23}F$, $^{22, 24, 25, 26}Ne$, $^{28, 29, 30}Mg$ and $^{32, 34}Si$ from various parents $^{199-226}Fr$, $^{206-232}Ac$, $^{209-237}Th$, $^{212-238}Pa$, $^{217-241}U$, $^{225-242}Np$, $^{225-244}Pu$, $^{231-246}Am$, $^{202-230}Ra$ and $^{233-249}Cm$ leading to doubly magic $^{208}Pb$ and neighbouring nuclei have been calculated by using the Coulomb and proximity potential model (CPPM). In CPPM, the external drifting potential barrier is obtained as the sum of the Coulomb potential, proximity potential and centrifugal potential for the touching configuration and for the separated fragments. The decay energy of the reaction is given as

$$Q = \Delta M - (\Delta M_1 + \Delta M_2) \tag{12}$$

and the possibility to have a cluster decay process is related to its exotermicity, $Q > 0$. Here $\Delta M, \Delta M_1, \Delta M_2$ are the mass excess of the parent, daughter and cluster respectively. The Q values for cluster decay are calculated using the experimental mass excess values of Audi *et al.*, [50].

The $T_{1/2}$ values for the respective cluster decays are also calculated using the Universal (UNIV) curve and the Universal decay law (UDL) for alpha and cluster decay modes and the Scaling Law of Horoi et al., for cluster decay and are compared with CPPM values. The cluster decay half lives calculated using CPPM, UNIV, UDL and the scaling law of Horoi and their comparisons are shown in figures 1-7. The plots for $\log_{10}(T_{1/2})$ against the neutron number of the daughter in the corresponding decay are given in these figures.

Several simple and effective relationships for the decay half lives are available with parameters which are obtained by fitting the experimental data. Among them the universal (UNIV) curves [51–54] which are derived by extending a fission theory to larger mass asymmetry should be mentioned with great importance. They are based on the quantum mechanical tunnelling process relationship [4, 55] of the disintegration constant $\lambda$, valid in both fission-like and $\alpha$-like theories and the partial decay half life $T$ of the parent nucleus is related to the disintegration constant $\lambda$ of the exponential decay law in time as

$$\lambda = \ln 2 / T = \nu S P_s \tag{13}$$

where $T$ is the half life and $\nu$, $S$, and $P_s$ are three model-dependent quantities: $\nu$ is the frequency of assaults on the barrier per second, $S$ is the pre-formation probability of the cluster at the nuclear surface (equal to the penetrability of the internal part of the barrier in a fission theory [51, 52]), and $P_s$ is the quantum penetrability of the external potential barrier. By using the decimal logarithm,

$$\log_{10} T(s) = -\log_{10} P - \log_{10} S + [\log_{10}(\ln 2) - \log_{10} \nu] \tag{14}$$

In order to derive the universal formula it was assumed that $\nu$ = constant and that $S$ depends only on the mass number of the emitted particle $A_e$ [52, 4]. A microscopic calculation of the pre-formation probability [56] of many clusters from $^8$Be to $^{46}$Ar had shown indeed that it is dependent only upon the size of the cluster. The corresponding numerical values [52] have been obtained by a fit with experimental data for $\alpha$ decay: $S_\alpha$ = 0.0143153, $\nu = 10^{22.01} s^{-1}$. The additive constant for an even-even nucleus

$$c_{ee} = [-\log_{10} \nu + \log_{10}(\ln 2)] = -22.16917 \tag{15}$$

and the decimal logarithm of the pre-formation factor

$$\log_{10} S = -0.598(A_e - 1) \tag{16}$$

The penetrability of an external Coulomb barrier, having separation distance at the touching configuration $R_a = R_t = R_d + R_e$ as the first turning point and the second turning point defined by $e^2 Z_d Z_e / R_b = Q$, may be found analytically as

$$-\log_{10} P_S = 0.22873(\mu_A Z_d Z_e R_b)^{1/2} \times [\arccos\sqrt{r} - \sqrt{r(1-r)}] \tag{17}$$

where $r = R_t / R_b$, $R_t = 1.2249(A_d^{1/3} + A_e^{1/3})$ and $R_b = 1.43998 Z_d Z_e / Q$

The liquid-drop-model radius constant $r_0$ = 1.2249fm and the mass tables [50] are used to calculate the released energy $Q$.

A new universal decay law (UDL) for $\alpha$-decay and cluster decay modes was introduced [42, 57] starting from $\alpha$-like (extension to the heavier cluster of $\alpha$-decay theory) R-matrix theory. Moreover, this UDL was presented in an interesting way, which makes it possible to represent on the same plot with a single straight line the logarithm of the half lives minus some quantity versus one of the two parameters ($\chi'$ and $\rho'$) that depend on the atomic and mass numbers of the daughter and emitted particles as well as the $Q$ value. The universal decay law was introduced starting from the microscopic mechanism of the charged-particle emission. The UDL relates the half-life of monopole radioactive decay with the $Q$ values of the outgoing particles as well as the masses and charges of the nuclei involved in the decay. The Universal Decay Law (UDL) can be written in the logarithmic form as

$$\log_{10}(T_{1/2}) = aZ_cZ_d\sqrt{\frac{A}{Q_c}} + b\sqrt{AZ_cZ_d(A_d^{1/3} + A_c^{1/3})} + c \tag{18}$$

$$= a\chi' + b\rho' + c \tag{19}$$

where a, b and c are the coefficient sets of eq. (19) that determined by fitting to experiments of both $\alpha$ and cluster decays [57], and are given as a = 0.4314, b = -0.4087 and c = -25.7725 The term $b\rho'+c$ includes the effects that induce the clusterization in the mother nucleus. This relation holds for the monopole radioactive decays of all clusters, and hence it is called the Universal Decay Law (UDL) [57].

A new empirical formula for cluster decay was introduced by Horoi et al [43], for determining the half lives of both the alpha and cluster decays and is given by the equation,

$$\log_{10} T_{1/2} = (a_1\mu^x + b_1)[(Z_1Z_2)^y/\sqrt{Q} - 7] + (a_2\mu^x + b_2) \tag{20}$$

where $\mu$ is the reduced mass. The six parameters are $a_1$ = 9.1, $b_1$ = -10.2, $a_2$ = 7.39, $b_2$ = -23.2, $x$ = 0.416 and $y$ = 0.613.

Fig 1 gives the plot for the cluster emission of $^{14}$C from $^{199-226}$Fr, $^{202-229}$Ra, $^{206-232}$Ac, and $^{209-234}$Th. The minima of the logarithmic half lives are found for the decay leading to the near doubly magic daughter $^{207}$Tl (Z = 81, N = 126), doubly magic $^{208}$Pb (Z = 82, N = 126), near doubly magic $^{209}$Bi (Z = 83, N = 126) and near doubly magic $^{210}$Po (Z = 84, N = 126) respectively. Fig 2 gives the plot for the cluster emission of $^{20}$O from $^{202-230}$Ra, $^{18}$O from $^{209-233}$Th, $^{20}$O from $^{209-235}$Th, and $^{22}$O from $^{209-237}$Th respectively. Here the minima of the logarithmic half lives are found for the decay leading to the near doubly magic daughter $^{206}$Hg (Z = 80, N = 126) for the cluster emission of $^{20}$O from $^{226}$Ra and for the decay leading to the doubly magic $^{208}$Pb (Z = 82, N = 126) for the cluster emission of $^{18}$O from $^{226}$Th, $^{20}$O from $^{228}$Th, and $^{22}$O from $^{230}$Th respectively. Fig 3 gives the plot for the cluster emission of $^{20}$O, $^{22}$Ne, $^{24}$Ne, and $^{26}$Ne respectively from $^{217-240}$U, $^{217-237}$U, $^{217-239}$U and $^{217-241}$U. The minima of the logarithmic half lives are found for the decay leading to near doubly magic daughter $^{210}$Po (Z = 84, N = 126) for the cluster emission of $^{20}$O from $^{230}$U, near doubly magic daughter $^{207}$Pb (Z = 82, N = 125) for the cluster emission of $^{22}$Ne from $^{229}$U and for the decay leading to doubly magic daughter $^{208}$Pb (Z = 82, N = 126) for the cluster emission of $^{24}$Ne and $^{26}$Ne from $^{232}$U and $^{234}$U respectively. Fig 4 gives the plot for the cluster emission of $^{24}$Ne from $^{209-234}$Th, $^{24}$Ne from $^{212-236}$Pa, $^{26}$Ne from $^{209-236}$Th, and $^{28}$Mg from $^{217-238}$U respectively. Here the minima of the logarithmic half lives are found for the decay leading to the near doubly magic daughter $^{205}$Hg (Z = 80, N = 125) for the cluster emission of $^{24}$Ne from $^{229}$Th, near doubly magic $^{207}$Tl (Z = 81, N = 126) for the cluster emission of $^{24}$Ne from $^{231}$Pa, near doubly magic $^{206}$Hg (Z = 80, N = 126) for the cluster emission of $^{26}$Ne from $^{232}$Th and near doubly magic $^{204}$Hg (Z = 80, N = 124) for the cluster emission of $^{28}$Mg from $^{232}$U respectively.

Fig 5 gives the plot for the cluster emission of $^{30}$Mg from $^{217-240}$U, $^{30}$Mg from $^{225-242}$Np $^{28}$Mg from $^{228-243}$Pu and $^{30}$Mg from $^{228-245}$Pu respectively. Here the minima of the logarithmic half lives are found for the decay leading to the near doubly magic daughter $^{206}$Hg (Z = 80, N = 126), near doubly magic $^{207}$Tl (Z = 81, N = 126), doubly magic $^{208}$Pb (Z = 82, N = 126) for the cluster emission of $^{28}$Mg from $^{236}$Pu and for the cluster emission of $^{30}$Mg from $^{238}$Pu respectively. Fig 6 gives the plot for the cluster emission of $^{32}$Si from $^{228-242}$Pu, $^{34}$Si from $^{233-249}$Cm, $^{34}$Si from $^{228-244}$Pu and $^{34}$Si from $^{231-246}$Am respectively. The minima of the logarithmic half lives are found for the decay leading to the near doubly magic $^{204}$Hg

(Z = 80, N = 124), doubly magic $^{208}$Pb (Z = 82, N = 126), near doubly magic $^{206}$Hg (Z = 80, N = 126) and near doubly magic $^{207}$Tl (Z = 81, N = 126), daughters respectively. Fig 7 gives the plot for the cluster emission of the odd clusters $^{15}$N, $^{23}$F, $^{25}$Ne and $^{29}$Mg respectively from $^{206-230}$Ac, $^{212-238}$Pa, $^{217-240}$U and $^{217-239}$U. Here the minima of the logarithmic half lives are found for the decay leading to the doubly magic $^{208}$Pb (Z = 82, N = 126) for the cluster emission of $^{15}$N from $^{223}$Ac, $^{23}$F from $^{231}$Pa, $^{25}$Ne from $^{233}$U; and for the decay leading to the near doubly magic $^{206}$Hg (Z = 80, N = 126) for the cluster emission of $^{29}$Mg from $^{235}$U. Of these plots, it can be seen that plots for the emission of the odd clusters i.e. $^{25}$Ne from $^{217-240}$U and $^{29}$Mg $^{217-239}$U reveal the odd-even staggering (OES). The abrupt changes in binding energy as one goes from a nucleus with an even number of neutrons (or protons) to its neighbour with an odd number of nucleons are known as odd-even-stagger (OES). The odd-even-stagger (OES) in atomic nuclei is usually attributed to the existence of nucleonic pairing correlations [58, 59].

All the plots connecting $\log_{10}(T_{1/2})$ versus neutron number of daughter nuclei reveal that the four calculations CPPM, UNIV, UDL and Scaling law show the same trend. It should be noted that the CPPM values matches well with the UDL values than that of the UNIV or the values obtained using the scaling law of Horoi. The low values of the cluster decay half lives at N=126 reveal the role of neutron magicity. It can be seen that the $\log_{10}(T_{1/2})$ have the lowest value in those decays leading to the doubly magic daughter nucleus $^{208}$Pb (Z = 82, N = 126) and the near doubly magic daughter nuclei $^{206}$Hg (Z = 80, N = 126), $^{207}$Tl (Z = 81, N = 126), $^{209}$Bi (Z =83, N =126) and $^{210}$Po (Z =84, N = 126). Thus our study reveals the fact that in cluster radioactivity the role of neutron shell closure is crucial than proton shell closure.

Fig 8 represents the Geiger – Nuttal plots for $\log_{10}(T_{1/2})$ versus $Q^{-1/2}$ for the various clusters $^{14}$C, $^{15}$N, $^{18,20,22}$O, $^{23}$F, $^{22,24,25,26}$Ne, $^{28,29,30}$Mg and $^{32,34}$Si from the parents $^{199-226}$Fr,

$^{206\text{-}232}$Ac, $^{209\text{-}237}$Th, $^{212\text{-}238}$Pa, $^{217\text{-}241}$U, $^{225\text{-}242}$Np, $^{228\text{-}245}$Pu, $^{231\text{-}246}$Am and $^{233\text{-}249}$Cm. These plots are found to be linear with different slopes and intercepts. We would like to point out that the Geiger – Nuttal law is for pure Coulomb potential but our present study reveals that inclusion of proximity potential will not produce much deviation to the linear nature of these plots which agrees with our earlier observations. We would also like to mention that the presence of proximity potential (nuclear structure effect) and shell effect (through Q value) are evident from the observed variation in the slope and intercept of Geiger-Nuttall plots for different clusters from various parents.

In the Tables 1-6 give computed Q values, barrier penetrability, decay constant and half lives for the most probable cluster emissions $(T_{1/2} < 10^{30}\,s)$. The parent nuclei, the emitted clusters and the corresponding daughter nuclei are given in column 1, 2 and 3 respectively of the tables mentioned above. Column 4 gives the respective Q values of these decays which are evaluated using equation (12). The penetrability and decay constants for the respective decays are calculated using CPPM and are included in columns 5 and 6 respectively. The cluster decay half lives calculated using CPPM is arranged in column 7. The experimental cluster decay half lives are available only for a limited number of decays for each parent-cluster combinations. Those values that are available are given in column 8. A comparison of our calculated values with that of the experimental half lives reveals that the computed half-lives for cluster decay are in good agreement with the experimental data.

In order to convince the agreement with experimental data we have calculated the standard deviation of $\log_{10}(T_{1/2})$ values for the CPPM, and have compared it with the calculated standard deviation of $\log_{10}(T_{1/2})$ values of UNIV, UDL and Scaling law of Horoi and have tabulated in Table 7. The standard deviation is given by

$$\sigma = \left[\frac{1}{(n-1)}\sum_{i=1}^{n}\left[\log(T_{1/2}^{cal.}) - \log(T_{1/2}^{\exp.})\right]^2\right]^{1/2} \qquad (21)$$

## 4. Conclusion

Using the Coulomb and proximity potential model (CPPM), the penetrability, decay constant and cluster decay half lives has been examined in detail for $^{199-226}$Fr, $^{206-232}$Ac, $^{209-237}$Th, $^{212-238}$Pa, $^{217-241}$U, $^{225-242}$Np, $^{228-245}$Pu, $^{231-246}$Am and $^{233-249}$Cm isotopes. All the parent-cluster combinations, where the experimental results were available are taken and we would confidently like to mention here that this is an elaborate theoretical study on cluster radioactivity done during the recent times. The results thus obtained were compared with the corresponding experimental data and with the values of UNIV, UDL and the scaling law of Horoi and it is found that they match well over a wide range. We have also performed the calculations for the emission of $^{22}$O, $^{20}$O, $^{20}$O from the parents $^{209-237}$Th, $^{202-230}$Ra and $^{217-240}$U respectively were the experimental values are not available and most of them are found to be most favourable for measurement $(T_{1/2} < 10^{30} s)$ and this observation also will serve as a guide to the future experiments. The odd-even staggering (OES) are found to be more prominent in the emission of odd mass clusters. Our study reveals the role of doubly magic $^{208}$Pb daughter nuclei in cluster decay process and also reveals the fact that the role of neutron shell closure is crucial than proton shell closure.

Table 1. Comparison of the logarithm of predicted cluster decay half lives with that of the experimental cluster half lives for the emission of the cluster $^{14}C$ from various isotopes of Fr, Ra, Ac and Th. The half lives are calculated for zero angular momentum transfers. $T_{1/2}$ is in seconds.

| Parent nuclei | Emitted cluster | Daughter nuclei | Q value (MeV) | Penetrability P | Decay constant $\lambda$ (s$^{-1}$) | $\log_{10}(T_{1/2})$ Expt. | $\log_{10}(T_{1/2})$ CPPM |
|---|---|---|---|---|---|---|---|
| $^{216}$Fr | $^{14}$C | $^{202}$Tl | 25.942 | 6.802x10$^{-50}$ | 4.839x10$^{-29}$ | | 28.156 |
| $^{217}$Fr | $^{14}$C | $^{203}$Tl | 27.056 | 1.302x10$^{-46}$ | 9.669x10$^{-26}$ | | 24.855 |
| $^{218}$Fr | $^{14}$C | $^{204}$Tl | 28.385 | 5.867x10$^{-43}$ | 4.568x10$^{-22}$ | | 21.181 |
| $^{219}$Fr | $^{14}$C | $^{205}$Tl | 29.418 | 2.836x10$^{-40}$ | 2.289x10$^{-19}$ | | 18.481 |
| $^{220}$Fr | $^{14}$C | $^{206}$Tl | 30.716 | 4.489x10$^{-37}$ | 3.782x10$^{-16}$ | | 15.263 |
| $^{221}$Fr | $^{14}$C | $^{207}$Tl | 31.292 | 1.062x10$^{-35}$ | 9.116x10$^{-15}$ | 14.52 | 13.881 |
| $^{222}$Fr | $^{14}$C | $^{208}$Tl | 30.078 | 1.672x10$^{-38}$ | 1.379x10$^{-17}$ | | 16.701 |
| $^{223}$Fr | $^{14}$C | $^{209}$Tl | 29.001 | 3.893x10$^{-41}$ | 3.096x10$^{-20}$ | | 19.349 |
| $^{224}$Fr | $^{14}$C | $^{210}$Tl | 27.886 | 5.427x10$^{-44}$ | 4.151x10$^{-23}$ | | 22.223 |
| $^{225}$Fr | $^{14}$C | $^{211}$Tl | 26.876 | 9.919x10$^{-47}$ | 7.313x10$^{-26}$ | | 24.977 |
| $^{226}$Fr | $^{14}$C | $^{212}$Tl | 26.000 | 3.116x10$^{-49}$ | 2.222x10$^{-28}$ | | 27.494 |
| | | | | | | | |
| $^{216}$Ra | $^{14}$C | $^{202}$Pb | 26.205 | 1.795x10$^{-50}$ | 1.290x10$^{-29}$ | | 28.730 |
| $^{217}$Ra | $^{14}$C | $^{203}$Pb | 27.648 | 2.655x10$^{-46}$ | 2.013x10$^{-25}$ | | 24.537 |
| $^{218}$Ra | $^{14}$C | $^{204}$Pb | 28.740 | 2.508x10$^{-43}$ | 1.977x10$^{-22}$ | | 21.545 |
| $^{219}$Ra | $^{14}$C | $^{205}$Pb | 30.144 | 9.808x10$^{-40}$ | 8.109x10$^{-19}$ | | 17.932 |
| $^{220}$Ra | $^{14}$C | $^{206}$Pb | 31.038 | 1.522x10$^{-37}$ | 1.295x10$^{-16}$ | | 15.728 |
| $^{221}$Ra | $^{14}$C | $^{207}$Pb | 32.395 | 2.116x10$^{-34}$ | 1.880x10$^{-13}$ | 13.39 | 12.567 |
| $^{222}$Ra | $^{14}$C | $^{208}$Pb | 33.049 | 6.388x10$^{-33}$ | 5.790x10$^{-12}$ | 11.01 | 11.078 |
| $^{223}$Ra | $^{14}$C | $^{209}$Pb | 31.828 | 1.438x10$^{-35}$ | 1.255x10$^{-14}$ | 15.20 | 13.742 |
| $^{224}$Ra | $^{14}$C | $^{210}$Pb | 30.535 | 1.560x10$^{-38}$ | 1.306x10$^{-17}$ | 15.68 | 16.725 |
| $^{225}$Ra | $^{14}$C | $^{211}$Pb | 29.465 | 4.099x10$^{-41}$ | 3.313x10$^{-20}$ | | 19.321 |
| $^{226}$Ra | $^{14}$C | $^{212}$Pb | 28.196 | 2.327x10$^{-44}$ | 1.800x10$^{-23}$ | 21.19 | 22.585 |
| $^{227}$Ra | $^{14}$C | $^{213}$Pb | 27.343 | 1.207x10$^{-46}$ | 9.053x10$^{-26}$ | | 24.884 |
| $^{228}$Ra | $^{14}$C | $^{214}$Pb | 26.102 | 3.536x10$^{-50}$ | 2.532x10$^{-29}$ | | 28.437 |
| $^{229}$Ra | $^{14}$C | $^{215}$Pb | 25.063 | 2.545x10$^{-53}$ | 1.749x10$^{-32}$ | | 31.598 |
| | | | | | | | |
| $^{216}$Ac | $^{14}$C | $^{202}$Bi | 25.836 | 6.051x10$^{-53}$ | 4.289x10$^{-32}$ | | 31.208 |
| $^{217}$Ac | $^{14}$C | $^{203}$Bi | 27.227 | 8.506x10$^{-49}$ | 6.353x10$^{-28}$ | | 27.038 |
| $^{218}$Ac | $^{14}$C | $^{204}$Bi | 28.487 | 2.820x10$^{-45}$ | 2.204x10$^{-24}$ | | 23.498 |
| $^{219}$Ac | $^{14}$C | $^{205}$Bi | 29.612 | 2.583x10$^{-42}$ | 2.097x10$^{-21}$ | | 20.519 |
| $^{220}$Ac | $^{14}$C | $^{206}$Bi | 30.760 | 1.987x10$^{-39}$ | 1.677x10$^{-18}$ | | 17.616 |
| $^{221}$Ac | $^{14}$C | $^{207}$Bi | 31.554 | 1.643x10$^{-37}$ | 1.422x10$^{-16}$ | | 15.688 |
| $^{222}$Ac | $^{14}$C | $^{208}$Bi | 32.471 | 2.295x10$^{-35}$ | 2.044x10$^{-14}$ | | 13.530 |
| $^{223}$Ac | $^{14}$C | $^{209}$Bi | 33.064 | 5.173x10$^{-34}$ | 4.691x10$^{-13}$ | 12.60 | 12.169 |
| $^{224}$Ac | $^{14}$C | $^{210}$Bi | 32.006 | 2.695x10$^{-36}$ | 2.366x10$^{-15}$ | | 14.467 |
| $^{225}$Ac | $^{14}$C | $^{211}$Bi | 30.476 | 7.708x10$^{-40}$ | 6.444x10$^{-19}$ | 17.16 | 18.032 |

Table 1 continued..

| Parent nuclei | Emitted cluster | Daughter nuclei | Q value (MeV) | Penetrability P | Decay constant $\lambda$ (s$^{-1}$) | log$_{10}$(T$_{1/2}$) Expt. | log$_{10}$(T$_{1/2}$) CPPM |
|---|---|---|---|---|---|---|---|
| $^{226}$Ac | $^{14}$C | $^{212}$Bi | 29.407 | 1.876x10$^{-42}$ | 1.513x10$^{-21}$ | | 20.661 |
| $^{227}$Ac | $^{14}$C | $^{213}$Bi | 28.062 | 5.816x10$^{-46}$ | 4.476x10$^{-25}$ | | 24.189 |
| $^{228}$Ac | $^{14}$C | $^{214}$Bi | 27.076 | 1.194x10$^{-48}$ | 8.874x10$^{-28}$ | | 26.893 |
| $^{229}$Ac | $^{14}$C | $^{215}$Bi | 26.081 | 1.535x10$^{-51}$ | 1.098x10$^{-30}$ | | 29.799 |
| $^{217}$Th | $^{14}$C | $^{203}$Po | 26.504 | 2.850x10$^{-52}$ | 2.071x10$^{-31}$ | | 30.524 |
| $^{218}$Th | $^{14}$C | $^{204}$Po | 27.689 | 8.947x10$^{-49}$ | 6.795x10$^{-28}$ | | 27.009 |
| $^{219}$Th | $^{14}$C | $^{205}$Po | 28.960 | 2.876x10$^{-45}$ | 2.284x10$^{-24}$ | | 23.482 |
| $^{220}$Th | $^{14}$C | $^{206}$Po | 29.832 | 5.766x10$^{-43}$ | 4.717x10$^{-22}$ | | 21.167 |
| $^{221}$Th | $^{14}$C | $^{207}$Po | 31.065 | 7.342x10$^{-40}$ | 6.257x10$^{-19}$ | | 18.044 |
| $^{222}$Th | $^{14}$C | $^{208}$Po | 31.653 | 1.958x10$^{-38}$ | 1.699x10$^{-17}$ | | 16.610 |
| $^{223}$Th | $^{14}$C | $^{209}$Po | 32.732 | 6.445x10$^{-36}$ | 5.785x10$^{-15}$ | | 14.078 |
| $^{224}$Th | $^{14}$C | $^{210}$Po | 32.930 | 2.032x10$^{-35}$ | 1.835x10$^{-14}$ | | 13.577 |
| $^{225}$Th | $^{14}$C | $^{211}$Po | 31.723 | 4.156x10$^{-38}$ | 3.615x10$^{-17}$ | | 16.283 |
| $^{226}$Th | $^{14}$C | $^{212}$Po | 30.547 | 7.596x10$^{-41}$ | 6.365x10$^{-20}$ | | 19.037 |
| $^{227}$Th | $^{14}$C | $^{213}$Po | 29.440 | 1.399x10$^{-43}$ | 1.130x10$^{-22}$ | | 21.788 |
| $^{228}$Th | $^{14}$C | $^{214}$Po | 28.222 | 9.165x10$^{-47}$ | 7.094x10$^{-26}$ | | 24.989 |
| $^{229}$Th | $^{14}$C | $^{215}$Po | 27.107 | 7.852x10$^{-50}$ | 5.838x10$^{-29}$ | | 28.074 |
| $^{230}$Th | $^{14}$C | $^{216}$Po | 26.061 | 6.695x10$^{-53}$ | 4.786x10$^{-32}$ | | 31.161 |

Table 2. Comparison of the logarithm of predicted cluster decay half lives with that of the experimental cluster half lives for the emission of the clusters $^{18, 20, 22}$O from various isotopes of Ra, Th and U. The half lives are calculated for zero angular momentum transfers. $T_{1/2}$ is in seconds.

| Parent nuclei | Emitted cluster | Daughter nuclei | Q value (MeV) | Penetrability P | Decay constant $\lambda$ (s$^{-1}$) | $\log_{10}(T_{1/2})$ Expt. | $\log_{10}(T_{1/2})$ CPPM |
|---|---|---|---|---|---|---|---|
| $^{223}$Ra | $^{20}$O | $^{203}$Hg | 38.706 | 5.590x10$^{-55}$ | 5.87x10$^{-34}$ | | 33.072 |
| $^{224}$Ra | $^{20}$O | $^{204}$Hg | 39.719 | 2.533x10$^{-52}$ | 2.73x10$^{-31}$ | | 30.404 |
| $^{225}$Ra | $^{20}$O | $^{205}$Hg | 40.483 | 2.310x10$^{-50}$ | 2.54x10$^{-29}$ | | 28.441 |
| $^{226}$Ra | $^{20}$O | $^{206}$Hg | 40.817 | 1.746x10$^{-49}$ | 1.93x10$^{-28}$ | | 27.550 |
| $^{227}$Ra | $^{20}$O | $^{207}$Hg | 39.601 | 1.984x10$^{-52}$ | 2.13x10$^{-31}$ | | 30.512 |
| $^{228}$Ra | $^{20}$O | $^{208}$Hg | 38.254 | 7.588x10$^{-56}$ | 7.87x10$^{-35}$ | | 33.940 |
| $^{219}$Th | $^{18}$O | $^{201}$Pb | 40.509 | 1.589x10$^{-53}$ | 1.747x10$^{-32}$ | | 31.598 |
| $^{220}$Th | $^{18}$O | $^{202}$Pb | 41.384 | 2.097x10$^{-51}$ | 2.357x10$^{-30}$ | | 29.468 |
| $^{221}$Th | $^{18}$O | $^{203}$Pb | 42.506 | 8.647x10$^{-49}$ | 9.980x10$^{-28}$ | | 26.842 |
| $^{222}$Th | $^{18}$O | $^{204}$Pb | 43.093 | 2.026x10$^{-47}$ | 2.371x10$^{-26}$ | | 25.466 |
| $^{223}$Th | $^{18}$O | $^{205}$Pb | 43.937 | 1.596x10$^{-45}$ | 1.904x10$^{-24}$ | | 23.561 |
| $^{224}$Th | $^{18}$O | $^{206}$Pb | 44.562 | 3.954x10$^{-44}$ | 4.784x10$^{-23}$ | | 22.161 |
| $^{225}$Th | $^{18}$O | $^{207}$Pb | 45.542 | 4.944x10$^{-42}$ | 6.114x10$^{-21}$ | | 20.054 |
| $^{226}$Th | $^{18}$O | $^{208}$Pb | 45.726 | 1.379x10$^{-41}$ | 1.712x10$^{-20}$ | >15.30 | 19.607 |
| $^{227}$Th | $^{18}$O | $^{209}$Pb | 44.201 | 1.122x10$^{-44}$ | 1.346x10$^{-23}$ | | 22.711 |
| $^{228}$Th | $^{18}$O | $^{210}$Pb | 42.281 | 8.654x10$^{-49}$ | 9.935x10$^{-28}$ | | 26.844 |
| $^{229}$Th | $^{18}$O | $^{211}$Pb | 40.858 | 5.528x10$^{-52}$ | 6.133x10$^{-31}$ | | 30.053 |
| $^{224}$Th | $^{20}$O | $^{204}$Pb | 41.307 | 8.228x10$^{-52}$ | 9.215x10$^{-31}$ | | 29.876 |
| $^{225}$Th | $^{20}$O | $^{205}$Pb | 42.282 | 2.121x10$^{-49}$ | 2.431x10$^{-28}$ | | 27.455 |
| $^{226}$Th | $^{20}$O | $^{206}$Pb | 43.184 | 3.162x10$^{-47}$ | 3.702x10$^{-26}$ | | 25.272 |
| $^{227}$Th | $^{20}$O | $^{207}$Pb | 44.459 | 2.802x10$^{-44}$ | 3.378x10$^{-23}$ | | 22.312 |
| $^{228}$Th | $^{20}$O | $^{208}$Pb | 44.722 | 1.253x10$^{-43}$ | 1.519x10$^{-22}$ | 20.72 | 21.659 |
| $^{229}$Th | $^{20}$O | $^{209}$Pb | 43.402 | 1.615x10$^{-46}$ | 1.900x10$^{-25}$ | | 24.562 |
| $^{230}$Th | $^{20}$O | $^{210}$Pb | 41.794 | 3.235x10$^{-50}$ | 3.666x10$^{-29}$ | | 28.277 |
| $^{231}$Th | $^{20}$O | $^{211}$Pb | 40.510 | 2.687x10$^{-53}$ | 2.952x10$^{-32}$ | | 31.371 |
| $^{227}$Th | $^{22}$O | $^{205}$Pb | 40.296 | 1.413x10$^{-54}$ | 1.54x10$^{-33}$ | | 32.994 |
| $^{228}$Th | $^{22}$O | $^{206}$Pb | 41.277 | 5.806x10$^{-52}$ | 6.49x10$^{-31}$ | | 30.032 |
| $^{229}$Th | $^{22}$O | $^{207}$Pb | 42.758 | 3.355x10$^{-48}$ | 3.88x10$^{-27}$ | | 26.251 |
| $^{230}$Th | $^{22}$O | $^{208}$Pb | 43.332 | 9.571x10$^{-47}$ | 1.12x10$^{-25}$ | | 24.794 |
| $^{231}$Th | $^{22}$O | $^{209}$Pb | 42.151 | 1.470x10$^{-49}$ | 1.68x10$^{-28}$ | | 27.623 |
| $^{232}$Th | $^{22}$O | $^{210}$Pb | 40.896 | 1.124x10$^{-52}$ | 1.25x10$^{-31}$ | | 30.752 |
| $^{233}$Th | $^{22}$O | $^{211}$Pb | 39.944 | 4.102x10$^{-55}$ | 4.44x10$^{-34}$ | | 33.193 |

Table 2 continued..

| Parent nuclei | Emitted cluster | Daughter nuclei | Q value (MeV) | Penetrability P | Decay constant $\lambda$ (s$^{-1}$) | $\log_{10}(T_{1/2})$ Expt. | $\log_{10}(T_{1/2})$ CPPM |
|---|---|---|---|---|---|---|---|
| $^{226}$U | $^{20}$O | $^{206}$Po | 41.713 | 4.556x10$^{-54}$ | 5.15x10$^{-33}$ | | 32.128 |
| $^{227}$U | $^{20}$O | $^{207}$Po | 42.370 | 2.141x10$^{-52}$ | 2.46x10$^{-31}$ | | 30.454 |
| $^{228}$U | $^{20}$O | $^{208}$Po | 42.897 | 4.590x10$^{-51}$ | 5.34x10$^{-30}$ | | 29.113 |
| $^{229}$U | $^{20}$O | $^{209}$Po | 43.779 | 6.251x10$^{-49}$ | 7.42x10$^{-28}$ | | 26.971 |
| $^{230}$U | $^{20}$O | $^{210}$Po | 43.770 | 7.071x10$^{-49}$ | 8.39x10$^{-28}$ | | 26.920 |
| $^{231}$U | $^{20}$O | $^{211}$Po | 42.442 | 6.285x10$^{-52}$ | 7.23x10$^{-31}$ | | 29.982 |
| $^{232}$U | $^{20}$O | $^{212}$Po | 41.181 | 6.100x10$^{-55}$ | 6.81x10$^{-34}$ | | 33.007 |

Table 3. Comparison of the logarithm of the predicted cluster decay half lives with that of the experimental cluster half lives for the emission of the clusters $^{22,24,26}$Ne from various isotopes of U, Th and Pa. The half lives are calculated for zero angular momentum transfers. $T_{1/2}$ is in seconds.

| Parent nuclei | Emitted cluster | Daughter nuclei | Q value (MeV) | Penetrability P | Decay constant $\lambda$ (s$^{-1}$) | $\log_{10}(T_{1/2})$ Expt. | $\log_{10}(T_{1/2})$ CPPM |
|---|---|---|---|---|---|---|---|
| $^{220}$U | $^{22}$Ne | $^{198}$Pb | 57.104 | 2.485x10$^{-54}$ | 3.846x10$^{-33}$ | | 32.256 |
| $^{221}$U | $^{22}$Ne | $^{199}$Pb | 57.842 | 9.368x10$^{-53}$ | 1.468x10$^{-31}$ | | 30.674 |
| $^{222}$U | $^{22}$Ne | $^{200}$Pb | 58.567 | 3.158x10$^{-51}$ | 5.012x10$^{-30}$ | | 29.141 |
| $^{223}$U | $^{22}$Ne | $^{201}$Pb | 59.122 | 4.774x10$^{-50}$ | 7.649x10$^{-29}$ | | 27.957 |
| $^{224}$U | $^{22}$Ne | $^{202}$Pb | 59.672 | 6.854x10$^{-49}$ | 1.108x10$^{-27}$ | | 26.796 |
| $^{225}$U | $^{22}$Ne | $^{203}$Pb | 60.188 | 8.249x10$^{-48}$ | 1.345x10$^{-26}$ | | 25.712 |
| $^{226}$U | $^{22}$Ne | $^{204}$Pb | 60.462 | 3.409x10$^{-47}$ | 5.585x10$^{-26}$ | | 25.094 |
| $^{227}$U | $^{22}$Ne | $^{205}$Pb | 60.816 | 1.968x10$^{-46}$ | 3.244x10$^{-25}$ | | 24.329 |
| $^{228}$U | $^{22}$Ne | $^{206}$Pb | 61.034 | 6.269x10$^{-46}$ | 1.036x10$^{-24}$ | | 23.825 |
| $^{229}$U | $^{22}$Ne | $^{207}$Pb | 61.686 | 1.253x10$^{-44}$ | 2.094x10$^{-23}$ | | 22.519 |
| $^{230}$U | $^{22}$Ne | $^{208}$Pb | 61.387 | 4.366x10$^{-45}$ | 7.263x10$^{-24}$ | 19.57 | 22.979 |
| $^{231}$U | $^{22}$Ne | $^{209}$Pb | 59.443 | 1.180x10$^{-48}$ | 1.901x10$^{-27}$ | | 26.562 |
| $^{232}$U | $^{22}$Ne | $^{210}$Pb | 57.362 | 1.221x10$^{-52}$ | 1.898x10$^{-31}$ | | 30.562 |
| $^{223}$U | $^{24}$Ne | $^{199}$Pb | 57.019 | 2.669x10$^{-54}$ | 4.123x10$^{-33}$ | | 32.226 |
| $^{224}$U | $^{24}$Ne | $^{200}$Pb | 57.905 | 2.421x10$^{-52}$ | 3.797x10$^{-31}$ | | 30.261 |
| $^{225}$U | $^{24}$Ne | $^{201}$Pb | 58.586 | 7.559x10$^{-51}$ | 1.199x10$^{-29}$ | | 28.762 |
| $^{226}$U | $^{24}$Ne | $^{202}$Pb | 59.214 | 1.788x10$^{-49}$ | 2.869x10$^{-28}$ | | 27.383 |
| $^{227}$U | $^{24}$Ne | $^{203}$Pb | 59.760 | 2.796x10$^{-48}$ | 4.527x10$^{-27}$ | | 26.185 |
| $^{228}$U | $^{24}$Ne | $^{204}$Pb | 60.285 | 3.860x10$^{-47}$ | 6.303x10$^{-26}$ | | 25.041 |
| $^{229}$U | $^{24}$Ne | $^{205}$Pb | 60.932 | 8.982x10$^{-46}$ | 1.482x10$^{-24}$ | | 23.669 |
| $^{230}$U | $^{24}$Ne | $^{206}$Pb | 61.351 | 7.298x10$^{-45}$ | 1.212x10$^{-23}$ | | 22.757 |
| $^{231}$U | $^{24}$Ne | $^{207}$Pb | 62.207 | 4.001x10$^{-43}$ | 6.742x10$^{-22}$ | | 21.012 |
| $^{232}$U | $^{24}$Ne | $^{208}$Pb | 62.309 | 7.754x10$^{-43}$ | 1.308x10$^{-21}$ | 21.08 | 20.724 |
| $^{233}$U | $^{24}$Ne | $^{209}$Pb | 60.485 | 2.816x10$^{-46}$ | 4.615x10$^{-25}$ | 24.83 | 24.177 |
| $^{234}$U | $^{24}$Ne | $^{210}$Pb | 58.825 | 1.650x10$^{-49}$ | 2.629x10$^{-28}$ | 25.92 | 27.421 |
| $^{235}$U | $^{24}$Ne | $^{211}$Pb | 57.362 | 1.925x10$^{-52}$ | 2.991x10$^{-31}$ | 27.42 | 30.365 |
| $^{236}$U | $^{24}$Ne | $^{212}$Pb | 55.944 | 2.253x10$^{-55}$ | 3.414x10$^{-34}$ | >25.9 | 33.307 |
| $^{230}$U | $^{26}$Ne | $^{204}$Pb | 56.294 | 3.487x10$^{-55}$ | 5.318x10$^{-34}$ | | 33.115 |
| $^{231}$U | $^{26}$Ne | $^{205}$Pb | 57.145 | 3.266x10$^{-53}$ | 5.055x10$^{-32}$ | | 31.137 |
| $^{232}$U | $^{26}$Ne | $^{206}$Pb | 57.930 | 2.039x10$^{-51}$ | 3.199x10$^{-30}$ | | 29.336 |
| $^{233}$U | $^{26}$Ne | $^{207}$Pb | 58.941 | 3.587x10$^{-49}$ | 5.727x10$^{-28}$ | | 27.083 |
| $^{234}$U | $^{26}$Ne | $^{208}$Pb | 59.464 | 5.525x10$^{-48}$ | 8.899x10$^{-27}$ | 25.92 | 25.891 |
| $^{235}$U | $^{26}$Ne | $^{209}$Pb | 58.104 | 8.931x10$^{-51}$ | 1.405x10$^{-29}$ | | 28.623 |
| $^{236}$U | $^{26}$Ne | $^{210}$Pb | 56.744 | 1.188x10$^{-53}$ | 1.826x10$^{-32}$ | >25.90 | 31.579 |

Table 3 continued..

| Parent nuclei | Emitted cluster | Daughter nuclei | Q value (MeV) | Penetrability P | Decay constant $\lambda$ (s$^{-1}$) | $\log_{10}(T_{1/2})$ Expt. | $\log_{10}(T_{1/2})$ CPPM |
|---|---|---|---|---|---|---|---|
| $^{223}$Th | $^{24}$Ne | $^{199}$Hg | 54.884 | 8.442x10$^{-55}$ | 1.255x10$^{-33}$ | | 32.742 |
| $^{224}$Th | $^{24}$Ne | $^{200}$Hg | 55.451 | 1.767x10$^{-53}$ | 2.655x10$^{-32}$ | | 31.417 |
| $^{225}$Th | $^{24}$Ne | $^{201}$Hg | 55.924 | 2.191x10$^{-52}$ | 3.319x10$^{-31}$ | | 30.319 |
| $^{226}$Th | $^{24}$Ne | $^{202}$Hg | 56.494 | 4.286x10$^{-51}$ | 6.559x10$^{-30}$ | | 29.024 |
| $^{227}$Th | $^{24}$Ne | $^{203}$Hg | 57.026 | 7.014x10$^{-50}$ | 1.083x10$^{-28}$ | | 27.806 |
| $^{228}$Th | $^{24}$Ne | $^{204}$Hg | 57.413 | 5.226x10$^{-49}$ | 8.128x10$^{-28}$ | | 26.931 |
| $^{229}$Th | $^{24}$Ne | $^{205}$Hg | 57.824 | 4.407x10$^{-48}$ | 6.903x10$^{-27}$ | | 26.002 |
| $^{230}$Th | $^{24}$Ne | $^{206}$Hg | 57.761 | 4.051x10$^{-48}$ | 6.339x10$^{-27}$ | 24.61 | 26.039 |
| $^{231}$Th | $^{24}$Ne | $^{207}$Hg | 55.988 | 1.030x10$^{-51}$ | 1.562x10$^{-30}$ | | 29.647 |
| $^{232}$Th | $^{24}$Ne | $^{208}$Hg | 54.509 | 8.121x10$^{-55}$ | 1.199x10$^{-33}$ | >29.20 | 32.762 |
| | | | | | | | |
| $^{222}$Pa | $^{24}$Ne | $^{198}$Tl | 55.561 | 1.787x10$^{-55}$ | 2.690x10$^{-34}$ | | 33.411 |
| $^{223}$Pa | $^{24}$Ne | $^{199}$Tl | 56.330 | 9.915x10$^{-54}$ | 1.513x10$^{-32}$ | | 31.661 |
| $^{224}$Pa | $^{24}$Ne | $^{200}$Tl | 56.869 | 1.708x10$^{-52}$ | 2.632x10$^{-31}$ | | 30.420 |
| $^{225}$Pa | $^{24}$Ne | $^{201}$Tl | 57.473 | 3.793x10$^{-51}$ | 5.905x10$^{-30}$ | | 29.069 |
| $^{226}$Pa | $^{24}$Ne | $^{202}$Tl | 57.967 | 5.064x10$^{-50}$ | 7.952x10$^{-29}$ | | 27.940 |
| $^{227}$Pa | $^{24}$Ne | $^{203}$Tl | 58.544 | 9.122x10$^{-49}$ | 1.446x10$^{-27}$ | | 26.680 |
| $^{228}$Pa | $^{24}$Ne | $^{204}$Tl | 59.221 | 2.639x10$^{-47}$ | 4.233x10$^{-26}$ | | 25.214 |
| $^{229}$Pa | $^{24}$Ne | $^{205}$Tl | 59.670 | 2.566x10$^{-46}$ | 4.147x10$^{-25}$ | | 24.223 |
| $^{230}$Pa | $^{24}$Ne | $^{206}$Tl | 60.379 | 7.902x10$^{-45}$ | 1.292x10$^{-23}$ | | 22.729 |
| $^{231}$Pa | $^{24}$Ne | $^{207}$Tl | 60.410 | 1.111x10$^{-44}$ | 1.819x10$^{-23}$ | 23.23 | 22.581 |
| $^{232}$Pa | $^{24}$Ne | $^{208}$Tl | 58.649 | 4.310x10$^{-48}$ | 6.848x10$^{-27}$ | | 26.005 |
| $^{233}$Pa | $^{24}$Ne | $^{209}$Tl | 57.079 | 3.114x10$^{-51}$ | 4.815x10$^{-30}$ | | 29.158 |
| $^{234}$Pa | $^{24}$Ne | $^{210}$Tl | 55.538 | 2.027x10$^{-54}$ | 3.051x10$^{-33}$ | | 32.356 |
| $^{235}$Pa | $^{24}$Ne | $^{211}$Tl | 54.367 | 6.746x10$^{-57}$ | 9.936x10$^{-36}$ | | 34.844 |
| $^{236}$Pa | $^{24}$Ne | $^{212}$Tl | 52.951 | 2.220x10$^{-53}$ | 3.365x10$^{-32}$ | | 31.314 |
| | | | | | | | |
| $^{229}$Th | $^{26}$Ne | $^{203}$Hg | 54.425 | 2.632x10$^{-55}$ | 3.880x10$^{-34}$ | | 33.252 |
| $^{230}$Th | $^{26}$Ne | $^{204}$Hg | 55.124 | 1.163x10$^{-53}$ | 1.737x10$^{-32}$ | | 31.601 |
| $^{231}$Th | $^{26}$Ne | $^{205}$Hg | 55.674 | 2.382x10$^{-52}$ | 3.592x10$^{-31}$ | | 30.285 |
| $^{232}$Th | $^{26}$Ne | $^{206}$Hg | 55.964 | 1.262x10$^{-51}$ | 1.912x10$^{-30}$ | >29.20 | 29.559 |
| $^{233}$Th | $^{26}$Ne | $^{207}$Hg | 54.523 | 9.059x10$^{-55}$ | 1.337x10$^{-33}$ | | 32.714 |

Table 4. Comparison of the logarithm of the predicted cluster decay half lives with that of the experimental cluster half lives for the emission of the clusters $^{28,30}$Mg from various isotopes of U, Np and Pu. The half lives are calculated for zero angular momentum transfers. $T_{1/2}$ is in seconds.

| Parent nuclei | Emitted cluster | Daughter nuclei | Q value (MeV) | Penetrability P | Decay constant $\lambda$ (s$^{-1}$) | $\log_{10}(T_{1/2})$ Expt. | $\log_{10}(T_{1/2})$ CPPM |
|---|---|---|---|---|---|---|---|
| $^{223}$U | $^{28}$Mg | $^{195}$Hg | 71.858 | 2.296x10$^{-55}$ | 4.469x10$^{-34}$ | | 33.191 |
| $^{224}$U | $^{28}$Mg | $^{196}$Hg | 72.559 | 6.828x10$^{-54}$ | 1.341x10$^{-32}$ | | 31.713 |
| $^{225}$U | $^{28}$Mg | $^{197}$Hg | 72.936 | 4.857x10$^{-53}$ | 9.595x10$^{-32}$ | | 30.859 |
| $^{226}$U | $^{28}$Mg | $^{198}$Hg | 73.302 | 3.125x10$^{-52}$ | 6.205x10$^{-31}$ | | 30.048 |
| $^{227}$U | $^{28}$Mg | $^{199}$Hg | 73.587 | 1.467x10$^{-51}$ | 2.924x10$^{-30}$ | | 29.375 |
| $^{228}$U | $^{28}$Mg | $^{200}$Hg | 73.747 | 3.886x10$^{-51}$ | 7.763x10$^{-30}$ | | 28.951 |
| $^{229}$U | $^{28}$Mg | $^{201}$Hg | 73.892 | 9.383x10$^{-51}$ | 1.877x10$^{-29}$ | | 28.567 |
| $^{230}$U | $^{28}$Mg | $^{202}$Hg | 73.979 | 1.814x10$^{-50}$ | 3.635x10$^{-29}$ | | 28.280 |
| $^{231}$U | $^{28}$Mg | $^{203}$Hg | 74.092 | 3.804x10$^{-50}$ | 7.635x10$^{-29}$ | | 27.958 |
| $^{232}$U | $^{28}$Mg | $^{204}$Hg | 74.318 | 1.330x10$^{-49}$ | 2.678x10$^{-28}$ | >22.26 | 27.413 |
| $^{233}$U | $^{28}$Mg | $^{205}$Hg | 74.225 | 1.165x10$^{-49}$ | 2.344x10$^{-28}$ | >27.59 | 27.471 |
| $^{234}$U | $^{28}$Mg | $^{206}$Hg | 74.110 | 8.921x10$^{-50}$ | 1.790x10$^{-28}$ | 27.54 | 27.588 |
| $^{235}$U | $^{28}$Mg | $^{207}$Hg | 72.158 | 2.219x10$^{-53}$ | 4.337x10$^{-32}$ | >28.10 | 31.204 |
| $^{236}$U | $^{28}$Mg | $^{208}$Hg | 70.564 | 2.112x10$^{-56}$ | 4.036x10$^{-35}$ | 27.58 | 34.235 |
| $^{232}$U | $^{30}$Mg | $^{202}$Hg | 70.866 | 6.098x10$^{-56}$ | 1.170x10$^{-34}$ | | 33.772 |
| $^{233}$U | $^{30}$Mg | $^{203}$Hg | 71.100 | 2.309x10$^{-55}$ | 4.447x10$^{-34}$ | | 33.193 |
| $^{234}$U | $^{30}$Mg | $^{204}$Hg | 71.747 | 6.100x10$^{-54}$ | 1.185x10$^{-32}$ | | 31.767 |
| $^{235}$U | $^{30}$Mg | $^{205}$Hg | 72.118 | 4.312x10$^{-53}$ | 8.422x10$^{-32}$ | | 30.915 |
| $^{236}$U | $^{30}$Mg | $^{206}$Hg | 72.303 | 1.258x10$^{-52}$ | 2.464x10$^{-31}$ | 27.58 | 30.449 |
| $^{237}$U | $^{30}$Mg | $^{207}$Hg | 70.522 | 4.004x10$^{-56}$ | 7.648x10$^{-35}$ | | 33.957 |
| $^{232}$Np | $^{30}$Mg | $^{202}$Tl | 72.254 | 1.470x10$^{-55}$ | 2.876x10$^{-34}$ | | 33.382 |
| $^{233}$Np | $^{30}$Mg | $^{203}$Tl | 72.622 | 1.060x10$^{-54}$ | 2.085x10$^{-33}$ | | 32.521 |
| $^{234}$Np | $^{30}$Mg | $^{204}$Tl | 73.213 | 2.064x10$^{-53}$ | 4.092x10$^{-32}$ | | 31.229 |
| $^{235}$Np | $^{30}$Mg | $^{205}$Tl | 73.776 | 3.577x10$^{-52}$ | 7.148x10$^{-31}$ | | 29.987 |
| $^{236}$Np | $^{30}$Mg | $^{206}$Tl | 74.544 | 1.421x10$^{-50}$ | 2.869x10$^{-29}$ | | 28.383 |
| $^{237}$Np | $^{30}$Mg | $^{207}$Tl | 74.818 | 6.414x10$^{-50}$ | 1.299x10$^{-28}$ | >26.93 | 27.727 |
| $^{238}$Np | $^{30}$Mg | $^{208}$Tl | 73.116 | 3.770x10$^{-53}$ | 7.467x10$^{-32}$ | | 30.968 |
| $^{239}$Np | $^{30}$Mg | $^{209}$Tl | 71.861 | 1.422x10$^{-55}$ | 2.767x10$^{-34}$ | | 33.399 |
| $^{228}$Pu | $^{28}$Mg | $^{200}$Pb | 77.351 | 4.124x10$^{-49}$ | 8.641x10$^{-28}$ | | 26.904 |
| $^{229}$Pu | $^{28}$Mg | $^{201}$Pb | 77.676 | 2.176x10$^{-48}$ | 4.577x10$^{-27}$ | | 26.180 |
| $^{230}$Pu | $^{28}$Mg | $^{202}$Pb | 77.886 | 7.049x10$^{-48}$ | 1.487x10$^{-26}$ | | 25.668 |
| $^{231}$Pu | $^{28}$Mg | $^{203}$Pb | 78.090 | 2.212x10$^{-47}$ | 4.678x10$^{-26}$ | | 25.171 |
| $^{232}$Pu | $^{28}$Mg | $^{204}$Pb | 78.493 | 1.570x10$^{-46}$ | 3.338x10$^{-25}$ | | 24.317 |
| $^{233}$Pu | $^{28}$Mg | $^{205}$Pb | 78.838 | 8.681x10$^{-46}$ | 1.853x10$^{-24}$ | | 23.573 |

Table 4 continued..

| Parent nuclei | Emitted cluster | Daughter nuclei | Q value (MeV) | Penetrability P | Decay constant $\lambda$ (s$^{-1}$) | log$_{10}$(T$_{1/2}$) Expt. | log$_{10}$(T$_{1/2}$) CPPM |
|---|---|---|---|---|---|---|---|
| $^{234}$Pu | $^{28}$Mg | $^{206}$Pb | 79.153 | 4.199x10$^{-45}$ | 9.003x10$^{-24}$ | | 22.886 |
| $^{235}$Pu | $^{28}$Mg | $^{207}$Pb | 79.653 | 4.292x10$^{-44}$ | 9.259x10$^{-23}$ | | 21.874 |
| $^{236}$Pu | $^{28}$Mg | $^{208}$Pb | 79.669 | 6.011x10$^{-44}$ | 1.297x10$^{-22}$ | 21.67 | 21.728 |
| $^{237}$Pu | $^{28}$Mg | $^{209}$Pb | 77.725 | 2.556x10$^{-47}$ | 5.381x10$^{-26}$ | | 25.109 |
| $^{238}$Pu | $^{28}$Mg | $^{210}$Pb | 75.911 | 1.539x10$^{-50}$ | 3.164x10$^{-29}$ | 25.70 | 28.340 |
| $^{239}$Pu | $^{28}$Mg | $^{211}$Pb | 74.099 | 7.725x10$^{-54}$ | 1.550x10$^{-32}$ | | 31.650 |
| $^{232}$Pu | $^{30}$Mg | $^{202}$Pb | 73.211 | 4.870x10$^{-56}$ | 9.656x10$^{-35}$ | | 33.856 |
| $^{233}$Pu | $^{30}$Mg | $^{203}$Pb | 73.748 | 7.537x10$^{-55}$ | 1.505x10$^{-33}$ | | 32.663 |
| $^{234}$Pu | $^{30}$Mg | $^{204}$Pb | 74.370 | 1.675x10$^{-53}$ | 3.375x10$^{-32}$ | | 31.312 |
| $^{235}$Pu | $^{30}$Mg | $^{205}$Pb | 74.865 | 2.052x10$^{-52}$ | 4.161x10$^{-31}$ | | 30.222 |
| $^{236}$Pu | $^{30}$Mg | $^{206}$Pb | 75.598 | 7.141x10$^{-51}$ | 1.462x10$^{-29}$ | | 28.676 |
| $^{237}$Pu | $^{30}$Mg | $^{207}$Pb | 76.455 | 4.128x10$^{-49}$ | 8.548x10$^{-28}$ | | 26.909 |
| $^{238}$Pu | $^{30}$Mg | $^{208}$Pb | 76.823 | 2.709x10$^{-48}$ | 5.636x10$^{-27}$ | 25.70 | 26.089 |
| $^{239}$Pu | $^{30}$Mg | $^{209}$Pb | 75.114 | 1.802x10$^{-51}$ | 3.667x10$^{-30}$ | | 29.276 |
| $^{240}$Pu | $^{30}$Mg | $^{210}$Pb | 73.766 | 5.167x10$^{-54}$ | 1.032x10$^{-32}$ | | 31.827 |

Table 5. Comparison of the logarithm of the predicted cluster decay half lives with that of the experimental cluster half lives for the emission of the clusters $^{32,34}$Si from various isotopes of Pu, Cm and Am. The half lives are calculated for zero angular momentum transfers. $T_{1/2}$ is in seconds.

| Parent nuclei | Emitted cluster | Daughter nuclei | Q value (MeV) | Penetrability P | Decay constant $\lambda$ (s$^{-1}$) | $\log_{10}(T_{1/2})$ Expt. | $\log_{10}(T_{1/2})$ CPPM |
|---|---|---|---|---|---|---|---|
| $^{228}$Pu | $^{32}$Si | $^{196}$Hg | 91.997 | 4.991x10$^{-51}$ | 1.243x10$^{-29}$ | | 28.746 |
| $^{229}$Pu | $^{32}$Si | $^{197}$Hg | 92.021 | 8.157x10$^{-51}$ | 2.032x10$^{-29}$ | | 28.533 |
| $^{230}$Pu | $^{32}$Si | $^{198}$Hg | 91.969 | 9.583x10$^{-51}$ | 2.386x10$^{-29}$ | | 28.463 |
| $^{231}$Pu | $^{32}$Si | $^{199}$Hg | 91.913 | 1.073x10$^{-50}$ | 2.671x10$^{-29}$ | | 28.414 |
| $^{232}$Pu | $^{32}$Si | $^{200}$Hg | 91.951 | 1.790x10$^{-50}$ | 4.457x10$^{-29}$ | | 28.192 |
| $^{233}$Pu | $^{32}$Si | $^{201}$Hg | 91.794 | 1.318x10$^{-50}$ | 3.277x10$^{-29}$ | | 28.325 |
| $^{234}$Pu | $^{32}$Si | $^{202}$Hg | 91.776 | 1.771x10$^{-50}$ | 4.404x10$^{-29}$ | | 28.197 |
| $^{235}$Pu | $^{32}$Si | $^{203}$Hg | 91.534 | 8.914x10$^{-51}$ | 2.209x10$^{-29}$ | | 28.496 |
| $^{236}$Pu | $^{32}$Si | $^{204}$Hg | 91.673 | 2.177x10$^{-50}$ | 5.405x10$^{-29}$ | | 28.108 |
| $^{237}$Pu | $^{32}$Si | $^{205}$Hg | 91.461 | 1.270x10$^{-50}$ | 3.147x10$^{-29}$ | | 28.343 |
| $^{238}$Pu | $^{32}$Si | $^{206}$Hg | 91.191 | 5.751x10$^{-51}$ | 1.420x10$^{-29}$ | 25.27 | 28.688 |
| $^{239}$Pu | $^{32}$Si | $^{207}$Hg | 88.890 | 5.758x10$^{-55}$ | 1.386x10$^{-33}$ | | 32.699 |
| $^{233}$Cm | $^{34}$Si | $^{199}$Pb | 92.475 | 4.299x10$^{-54}$ | 1.077x10$^{-32}$ | | 31.809 |
| $^{234}$Cm | $^{34}$Si | $^{200}$Pb | 92.924 | 4.102x10$^{-53}$ | 1.032x10$^{-31}$ | | 30.827 |
| $^{235}$Cm | $^{34}$Si | $^{201}$Pb | 93.125 | 1.374x10$^{-52}$ | 3.466x10$^{-31}$ | | 30.301 |
| $^{236}$Cm | $^{34}$Si | $^{202}$Pb | 93.781 | 3.042x10$^{-51}$ | 7.728x10$^{-30}$ | | 28.953 |
| $^{237}$Cm | $^{34}$Si | $^{203}$Pb | 94.024 | 1.194x10$^{-50}$ | 3.041x10$^{-29}$ | | 28.358 |
| $^{238}$Cm | $^{34}$Si | $^{204}$Pb | 94.466 | 1.062x10$^{-49}$ | 2.717x10$^{-28}$ | | 27.407 |
| $^{239}$Cm | $^{34}$Si | $^{205}$Pb | 94.917 | 9.704x10$^{-49}$ | 2.495x10$^{-27}$ | | 26.444 |
| $^{240}$Cm | $^{34}$Si | $^{206}$Pb | 95.467 | 1.318x10$^{-47}$ | 3.409x10$^{-26}$ | | 25.308 |
| $^{241}$Cm | $^{34}$Si | $^{207}$Pb | 96.111 | 2.602x10$^{-46}$ | 6.774x10$^{-25}$ | | 24.009 |
| $^{242}$Cm | $^{34}$Si | $^{208}$Pb | 96.510 | 1.862x10$^{-45}$ | 4.867x10$^{-24}$ | 23.15 | 23.154 |
| $^{243}$Cm | $^{34}$Si | $^{209}$Pb | 94.754 | 1.869x10$^{-48}$ | 4.796x10$^{-27}$ | | 26.159 |
| $^{244}$Cm | $^{34}$Si | $^{210}$Pb | 93.138 | 3.012x10$^{-51}$ | 7.599x10$^{-30}$ | | 28.959 |
| $^{245}$Cm | $^{34}$Si | $^{211}$Pb | 91.452 | 3.238x10$^{-54}$ | 8.021x10$^{-33}$ | | 31.937 |
| $^{230}$Pu | $^{34}$Si | $^{196}$Hg | 88.717 | 5.844x10$^{-56}$ | 1.404x10$^{-34}$ | | 33.693 |
| $^{231}$Pu | $^{34}$Si | $^{197}$Hg | 88.783 | 1.081x10$^{-55}$ | 2.599x10$^{-34}$ | | 33.426 |
| $^{232}$Pu | $^{34}$Si | $^{198}$Hg | 89.277 | 1.313x10$^{-54}$ | 3.176x10$^{-33}$ | | 32.339 |
| $^{233}$Pu | $^{34}$Si | $^{199}$Hg | 89.554 | 5.894x10$^{-54}$ | 1.429x10$^{-32}$ | | 31.686 |
| $^{234}$Pu | $^{34}$Si | $^{200}$Hg | 89.811 | 2.513x10$^{-53}$ | 6.114x10$^{-32}$ | | 31.054 |
| $^{235}$Pu | $^{34}$Si | $^{201}$Hg | 89.804 | 3.361x10$^{-53}$ | 8.176x10$^{-32}$ | | 30.928 |
| $^{236}$Pu | $^{34}$Si | $^{202}$Hg | 90.205 | 2.681x10$^{-52}$ | 6.550x10$^{-31}$ | | 30.025 |
| $^{237}$Pu | $^{34}$Si | $^{203}$Hg | 90.319 | 5.911x10$^{-52}$ | 1.445x10$^{-30}$ | | 29.681 |
| $^{238}$Pu | $^{34}$Si | $^{204}$Hg | 90.811 | 6.508x10$^{-51}$ | 1.600x10$^{-29}$ | | 28.636 |
| $^{239}$Pu | $^{34}$Si | $^{205}$Hg | 90.833 | 9.675x10$^{-51}$ | 2.380x10$^{-29}$ | | 28.464 |

Table 5 continued..

| Parent nuclei | Emitted cluster | Daughter nuclei | Q value (MeV) | Penetrability P | Decay constant $\lambda$ (s$^{-1}$) | $\log_{10}(T_{1/2})$ Expt. | $\log_{10}(T_{1/2})$ CPPM |
|---|---|---|---|---|---|---|---|
| $^{240}$Pu | $^{34}$Si | $^{206}$Hg | 91.023 | 2.941x10$^{-50}$ | 7.250x10$^{-29}$ | >25.52 | 27.980 |
| $^{241}$Pu | $^{34}$Si | $^{207}$Hg | 89.133 | 1.224x10$^{-53}$ | 2.955x10$^{-32}$ | | 31.370 |
| $^{242}$Pu | $^{34}$Si | $^{208}$Hg | 87.775 | 4.632x10$^{-56}$ | 1.101x10$^{-34}$ | | 33.799 |
| $^{231}$Am | $^{34}$Si | $^{197}$Tl | 90.738 | 7.829x10$^{-55}$ | 1.924x10$^{-33}$ | | 32.227 |
| $^{232}$Am | $^{34}$Si | $^{198}$Tl | 90.847 | 1.806x10$^{-54}$ | 4.443x10$^{-33}$ | | 32.193 |
| $^{233}$Am | $^{34}$Si | $^{199}$Tl | 91.186 | 1.095x10$^{-53}$ | 2.705x10$^{-32}$ | | 31.409 |
| $^{234}$Am | $^{34}$Si | $^{200}$Tl | 91.535 | 6.868x10$^{-53}$ | 1.702x10$^{-31}$ | | 30.609 |
| $^{235}$Am | $^{34}$Si | $^{201}$Tl | 91.799 | 2.800x10$^{-52}$ | 6.960x10$^{-31}$ | | 29.998 |
| $^{236}$Am | $^{34}$Si | $^{202}$Tl | 92.12 | 1.586x10$^{-51}$ | 3.956x10$^{-30}$ | | 29.243 |
| $^{237}$Am | $^{34}$Si | $^{203}$Tl | 92.288 | 4.551x10$^{-51}$ | 1.137x10$^{-29}$ | | 28.785 |
| $^{238}$Am | $^{34}$Si | $^{204}$Tl | 92.723 | 3.859x10$^{-50}$ | 9.691x10$^{-29}$ | | 27.854 |
| $^{239}$Am | $^{34}$Si | $^{205}$Tl | 93.169 | 3.521x10$^{-49}$ | 8.885x10$^{-28}$ | | 26.892 |
| $^{240}$Am | $^{34}$Si | $^{206}$Tl | 93.722 | 4.820x10$^{-48}$ | 1.223x10$^{-26}$ | | 25.753 |
| $^{241}$Am | $^{34}$Si | $^{207}$Tl | 93.927 | 1.589x10$^{-47}$ | 4.043x10$^{-26}$ | >24.41 | 25.234 |
| $^{242}$Am | $^{34}$Si | $^{208}$Tl | 92.176 | 1.431x10$^{-50}$ | 3.573x10$^{-29}$ | | 28.288 |
| $^{243}$Am | $^{34}$Si | $^{209}$Tl | 90.771 | 4.873x10$^{-53}$ | 1.197x10$^{-31}$ | | 30.762 |
| $^{244}$Am | $^{34}$Si | $^{210}$Tl | 89.084 | 4.500x10$^{-56}$ | 1.085x10$^{-34}$ | | 33.805 |

Table 6. Comparison of the logarithm of the predicted cluster decay half lives with that of the experimental cluster half lives for the emission of the odd clusters $^{15}$N, $^{23}$F, $^{25}$Ne and $^{29}$Mg from various isotopes of Ac, Pa and U respectively. The half lives are calculated for zero angular momentum transfers. $T_{1/2}$ is in seconds.

| Parent nuclei | Emitted cluster | Daughter nuclei | Q value (MeV) | Penetrability P | Decay constant $\lambda$ (s$^{-1}$) | $\log_{10}(T_{1/2})$ Expt. | $\log_{10}(T_{1/2})$ CPPM |
|---|---|---|---|---|---|---|---|
| $^{206}$Ac | $^{15}$N | $^{191}$Pb | 33.658 | 3.900x10$^{-51}$ | 3.59x10$^{-30}$ | | 29.291 |
| $^{207}$Ac | $^{15}$N | $^{192}$Pb | 33.584 | 3.095x10$^{-51}$ | 2.89x10$^{-30}$ | | 29.393 |
| $^{208}$Ac | $^{15}$N | $^{193}$Pb | 32.848 | 5.221x10$^{-53}$ | 4.68x10$^{-32}$ | | 31.175 |
| $^{209}$Ac | $^{15}$N | $^{194}$Pb | 32.946 | 1.115x10$^{-52}$ | 1.00x10$^{-31}$ | | 30.842 |
| $^{210}$Ac | $^{15}$N | $^{195}$Pb | 32.402 | 5.287x10$^{-54}$ | 4.68x10$^{-33}$ | | 32.173 |
| $^{211}$Ac | $^{15}$N | $^{196}$Pb | 32.459 | 8.869x10$^{-54}$ | 7.86x10$^{-33}$ | | 31.950 |
| $^{212}$Ac | $^{15}$N | $^{197}$Pb | 31.927 | 4.194x10$^{-55}$ | 3.66x10$^{-34}$ | | 33.281 |
| $^{213}$Ac | $^{15}$N | $^{198}$Pb | 32.098 | 1.410x10$^{-54}$ | 1.24x10$^{-33}$ | | 32.748 |
| $^{214}$Ac | $^{15}$N | $^{199}$Pb | 31.555 | 5.874x10$^{-56}$ | 5.06x10$^{-35}$ | | 37.145 |
| $^{215}$Ac | $^{15}$N | $^{200}$Pb | 32.153 | 2.732x10$^{-54}$ | 2.40x10$^{-33}$ | | 32.460 |
| $^{216}$Ac | $^{15}$N | $^{201}$Pb | 33.279 | 2.542x10$^{-51}$ | 2.31x10$^{-30}$ | | 29.481 |
| $^{217}$Ac | $^{15}$N | $^{202}$Pb | 34.539 | 3.662x10$^{-48}$ | 3.45x10$^{-27}$ | | 26.302 |
| $^{218}$Ac | $^{15}$N | $^{203}$Pb | 35.525 | 8.811x10$^{-46}$ | 8.55x10$^{-25}$ | | 23.913 |
| $^{219}$Ac | $^{15}$N | $^{204}$Pb | 36.577 | 2.441x10$^{-43}$ | 2.44x10$^{-22}$ | | 21.455 |
| $^{220}$Ac | $^{15}$N | $^{205}$Pb | 37.420 | 1.964x10$^{-41}$ | 2.00x10$^{-20}$ | | 19.534 |
| $^{221}$Ac | $^{15}$N | $^{206}$Pb | 38.203 | 1.043x10$^{-39}$ | 1.09x10$^{-18}$ | | 17.803 |
| $^{222}$Ac | $^{15}$N | $^{207}$Pb | 38.971 | 4.666x10$^{-38}$ | 4.97x10$^{-17}$ | | 16.145 |
| $^{223}$Ac | $^{15}$N | $^{208}$Pb | 39.473 | 5.575x10$^{-37}$ | 6.01x10$^{-16}$ | >14.76 | 15.061 |
| $^{224}$Ac | $^{15}$N | $^{209}$Pb | 37.747 | 1.742x10$^{-40}$ | 1.79x10$^{-19}$ | | 18.586 |
| $^{225}$Ac | $^{15}$N | $^{210}$Pb | 36.264 | 1.148x10$^{-43}$ | 1.13x10$^{-22}$ | | 21.782 |
| $^{226}$Ac | $^{15}$N | $^{211}$Pb | 34.699 | 3.183x10$^{-47}$ | 3.02x10$^{-26}$ | | 25.360 |
| $^{227}$Ac | $^{15}$N | $^{212}$Pb | 33.296 | 1.336x10$^{-50}$ | 1.21x10$^{-29}$ | | 28.764 |
| $^{228}$Ac | $^{15}$N | $^{213}$Pb | 31.978 | 5.920x10$^{-54}$ | 5.15x10$^{-33}$ | | 32.132 |
| $^{227}$Pa | $^{23}$F | $^{204}$Pb | 48.611 | 8.034x10$^{-54}$ | 1.058x10$^{-32}$ | | 31.816 |
| $^{228}$Pa | $^{23}$F | $^{205}$Pb | 49.364 | 5.254x10$^{-52}$ | 7.026x10$^{-31}$ | | 29.994 |
| $^{229}$Pa | $^{23}$F | $^{206}$Pb | 50.353 | 1.066x10$^{-49}$ | 1.455x10$^{-28}$ | | 27.678 |
| $^{230}$Pa | $^{23}$F | $^{207}$Pb | 51.296 | 1.519x10$^{-47}$ | 2.111x10$^{-26}$ | | 25.516 |
| $^{231}$Pa | $^{23}$F | $^{208}$Pb | 51.843 | 2.769x10$^{-46}$ | 3.890x10$^{-25}$ | 26.02 | 24.251 |
| $^{232}$Pa | $^{23}$F | $^{209}$Pb | 50.232 | 9.902x10$^{-50}$ | 1.347x10$^{-28}$ | | 27.711 |
| $^{233}$Pa | $^{23}$F | $^{210}$Pb | 48.888 | 1.049x10$^{-52}$ | 1.389x10$^{-31}$ | | 30.698 |
| $^{234}$Pa | $^{23}$F | $^{211}$Pb | 47.502 | 6.841x10$^{-56}$ | 8.803x10$^{-35}$ | | 33.896 |

| Parent nuclei | Emitted cluster | Daughter nuclei | Q value (MeV) | Penetrability P | Decay constant $\lambda$ (s$^{-1}$) | $\log_{10}(T_{1/2})$ Expt. | $\log_{10}(T_{1/2})$ CPPM |
|---|---|---|---|---|---|---|---|
| $^{227}$U | $^{25}$Ne | $^{202}$Pb | 57.064 | 8.441x10$^{-54}$ | 1.30x10$^{-32}$ | | 31.725 |
| $^{228}$U | $^{25}$Ne | $^{203}$Pb | 56.120 | 9.507x10$^{-56}$ | 1.44x10$^{-34}$ | | 33.680 |
| $^{229}$U | $^{25}$Ne | $^{204}$Pb | 58.428 | 1.027x10$^{-50}$ | 1.63x10$^{-29}$ | | 28.629 |
| $^{230}$U | $^{25}$Ne | $^{205}$Pb | 57.493 | 1.362x10$^{-52}$ | 2.12x10$^{-31}$ | | 30.514 |
| $^{231}$U | $^{25}$Ne | $^{206}$Pb | 59.700 | 6.685x10$^{-48}$ | 1.09x10$^{-27}$ | | 25.806 |
| $^{232}$U | $^{25}$Ne | $^{207}$Pb | 59.169 | 6.774x10$^{-49}$ | 1.08x10$^{-26}$ | | 26.805 |
| $^{233}$U | $^{25}$Ne | $^{208}$Pb | 60.776 | 1.515x10$^{-45}$ | 2.49x10$^{-24}$ | 24.84 | 23.443 |
| $^{234}$U | $^{25}$Ne | $^{209}$Pb | 57.869 | 1.969x10$^{-51}$ | 3.09x10$^{-30}$ | | 29.351 |
| $^{235}$U | $^{25}$Ne | $^{210}$Pb | 57.756 | 1.389x10$^{-51}$ | 2.17x10$^{-30}$ | 27.42 | 29.503 |
| $^{229}$U | $^{29}$Mg | $^{200}$Hg | 71.334 | 1.669x10$^{-55}$ | 3.22x10$^{-34}$ | | 33.335 |
| $^{230}$U | $^{29}$Mg | $^{201}$Hg | 69.897 | 2.861x10$^{-58}$ | 5.42x10$^{-37}$ | | 36.116 |
| $^{231}$U | $^{29}$Mg | $^{202}$Hg | 71.771 | 2.089x10$^{-54}$ | 4.06x10$^{-33}$ | | 32.234 |
| $^{232}$U | $^{29}$Mg | $^{203}$Hg | 70.498 | 7.850x10$^{-57}$ | 1.50x10$^{-35}$ | | 34.672 |
| $^{233}$U | $^{29}$Mg | $^{204}$Hg | 72.229 | 2.771x10$^{-53}$ | 5.42x10$^{-23}$ | | 31.113 |
| $^{234}$U | $^{29}$Mg | $^{205}$Hg | 71.052 | 1.670x10$^{-55}$ | 3.21x10$^{-34}$ | | 33.335 |
| $^{235}$U | $^{29}$Mg | $^{206}$Hg | 72.485 | 1.444x10$^{-52}$ | 2.83x10$^{-31}$ | >28.09 | 30.390 |

Table 7. The comparisons of the standard deviation of the CPPM, UNIV, UDL and Horoi with the experimental cluster decay data.

| n | Parent | $\sigma_{CPPM}$ | $\sigma_{UNIV}$ | $\sigma_{UDL}$ | $\sigma_{HOROI}$ |
|---|---|---|---|---|---|
| 17 | even-even | 2.3888 | 1.0172 | 2.1067 | 1.0522 |
| 6 | even-odd | 1.9075 | 1.0962 | 1.5196 | 1.6289 |
| 5 | odd-even | 1.0861 | 0.9417 | 1.4355 | 1.5307 |
| 28 | all | 2.0584 | 0.9837 | 1.8671 | 1.2294 |

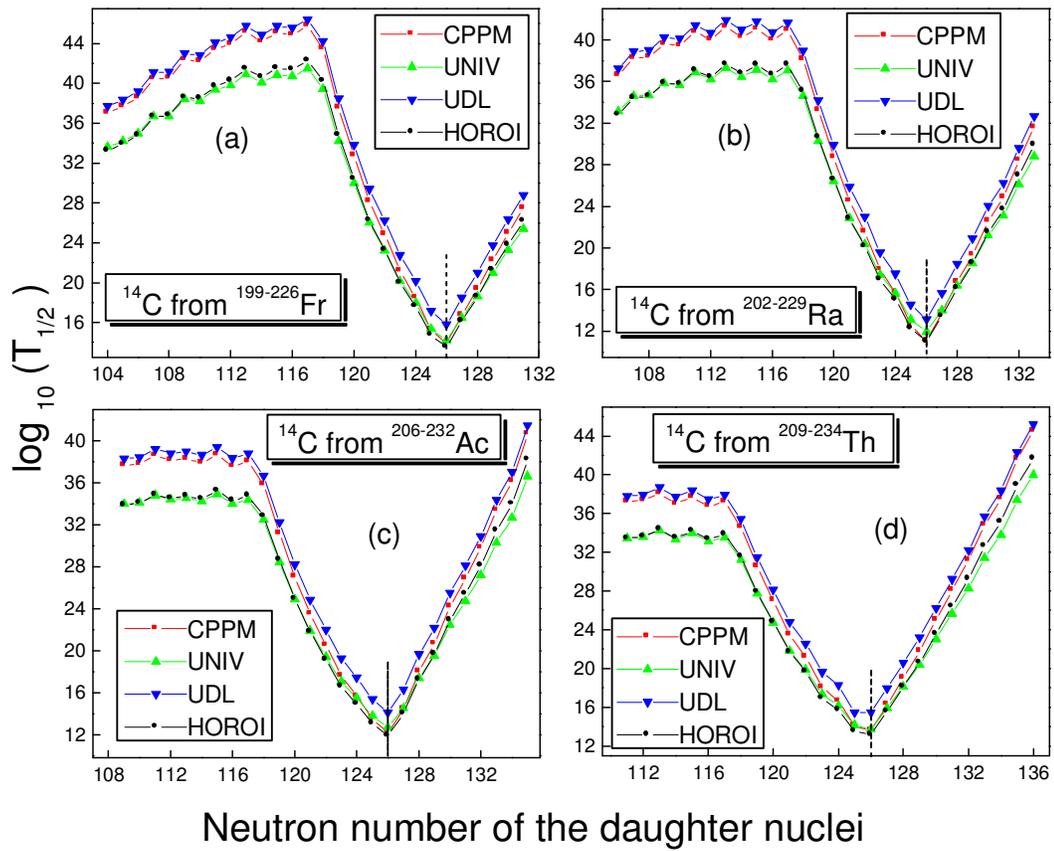

Fig.1. Plot of the computed $\log_{10}(T_{1/2})$ values vs. neutron number of daughter for the emission of cluster $^{14}C$ from Fr, Ra, Ac and Th isotopes. $T_{1/2}$ is in seconds.

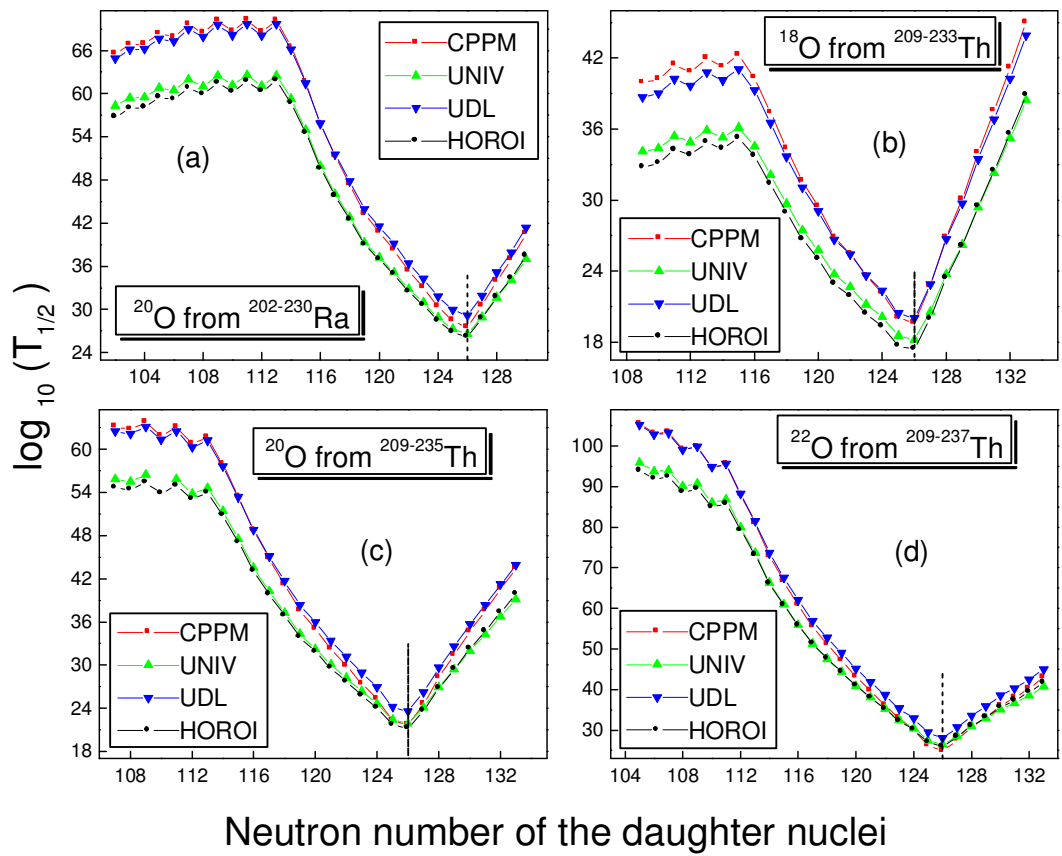

Fig.2. Plot of the computed $\log_{10}(T_{1/2})$ values vs. neutron number of daughter for the emission of cluster $^{20}$O from Ra and Th and clusters $^{18}$O and $^{22}$O from Th isotopes. $T_{1/2}$ is in seconds.

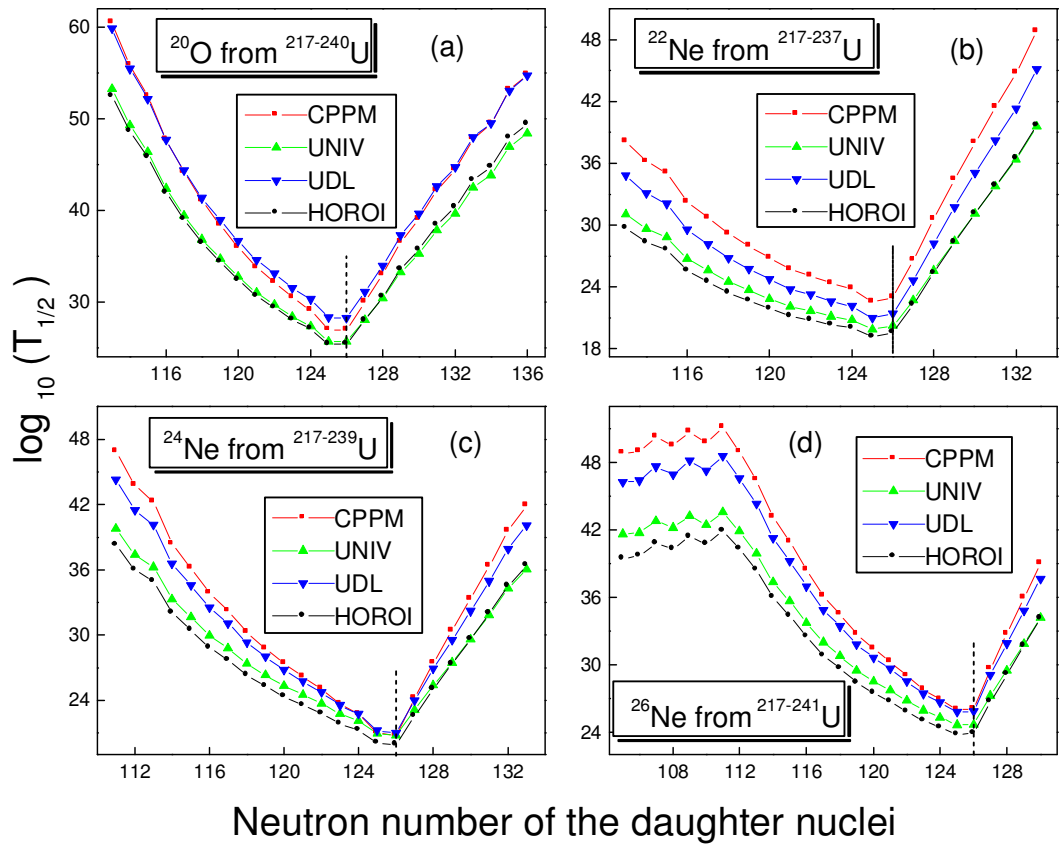

Fig.3. Plot of the computed $\log_{10}(T_{1/2})$ values vs. neutron number of daughter for the emission of clusters $^{20}$O, $^{22}$Ne, $^{24}$Ne and $^{26}$Ne from U isotopes. $T_{1/2}$ is in seconds.

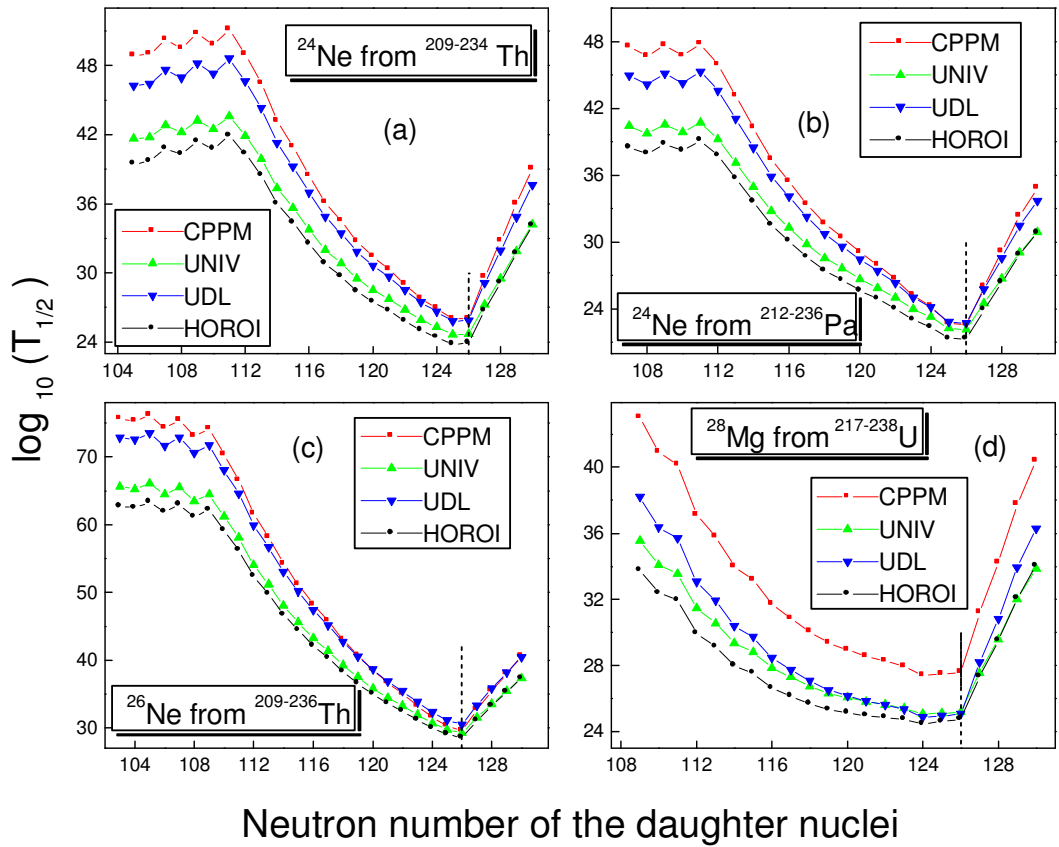

Fig.4. Plot of the computed $\log_{10}(T_{1/2})$ values vs. neutron number of daughter for the emission of clusters $^{24}$Ne from Th and Pa, $^{26}$Ne from Th and $^{28}$Mg from U isotopes. $T_{1/2}$ is in seconds.

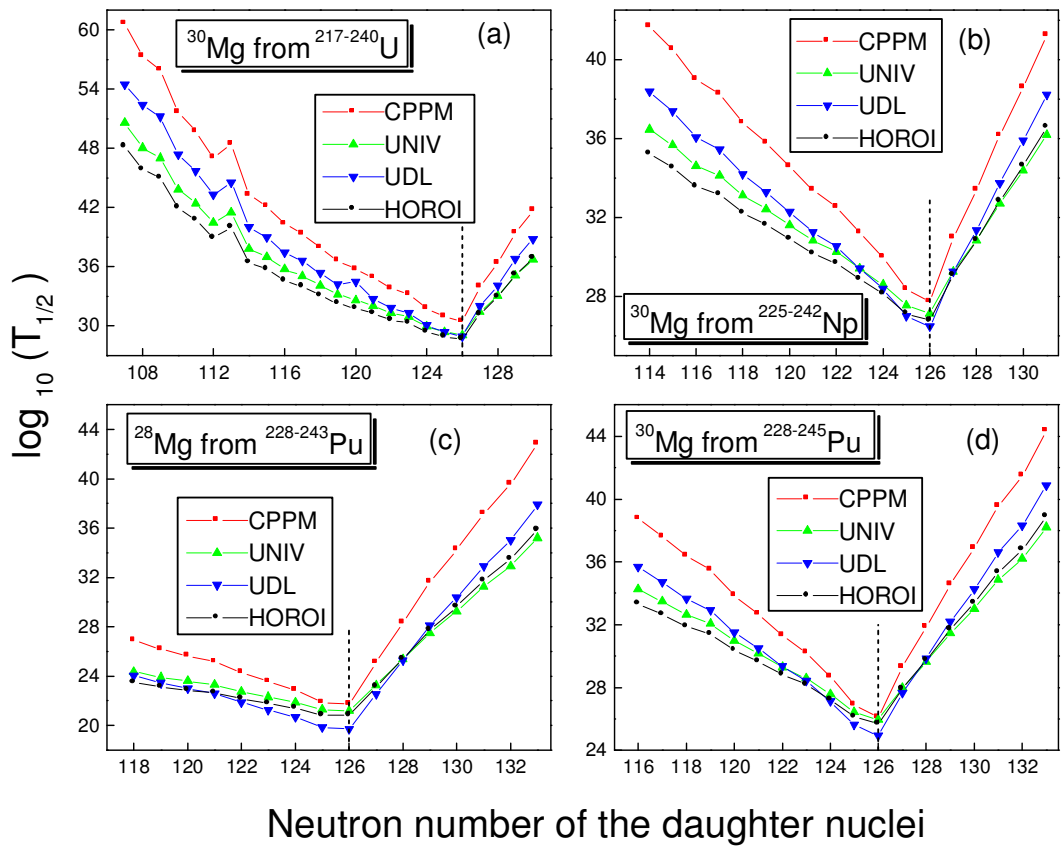

Fig.5. Plot of the computed $\log_{10}(T_{1/2})$ values vs. neutron number of daughter for the emission of clusters $^{28}$Mg from Pu and $^{30}$Mg from U, Np and Pu isotopes. $T_{1/2}$ is in seconds.

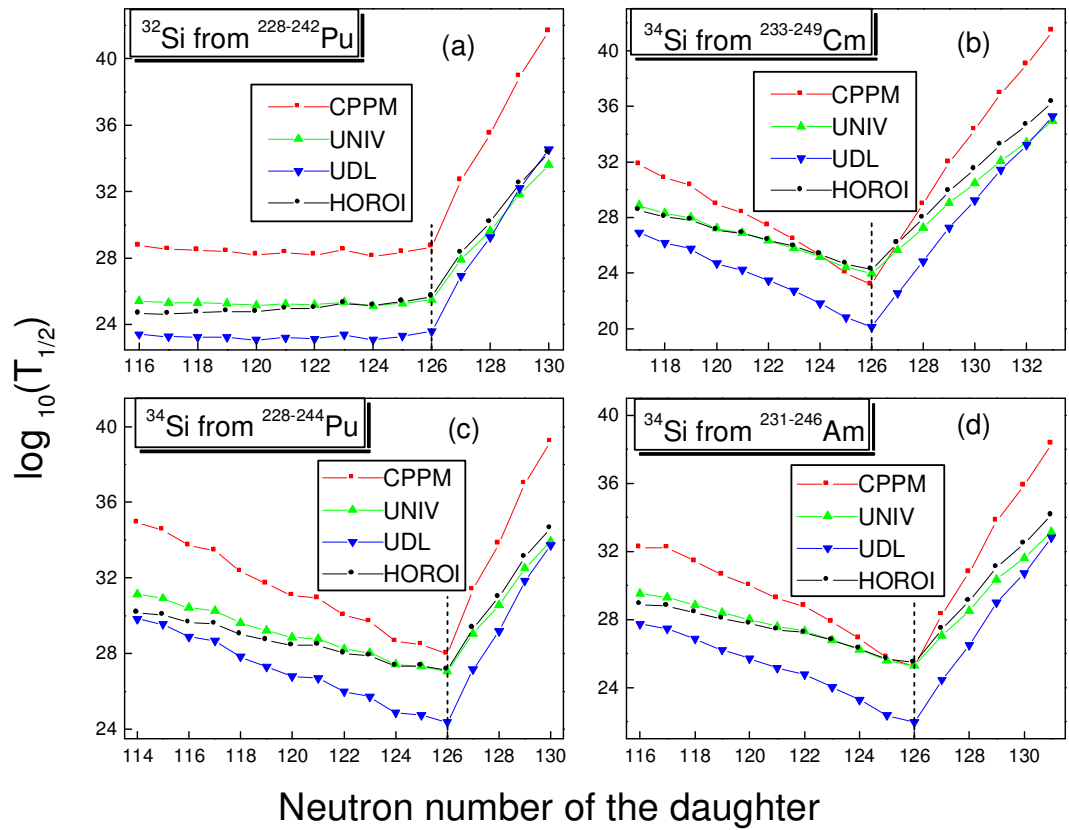

Fig.6. Plot of the computed $\log_{10}(T_{1/2})$ values vs. neutron number of daughter for the emission of clusters $^{32}$Si from Pu and $^{34}$Si from Cu, Pu and Am isotopes. $T_{1/2}$ is in seconds.

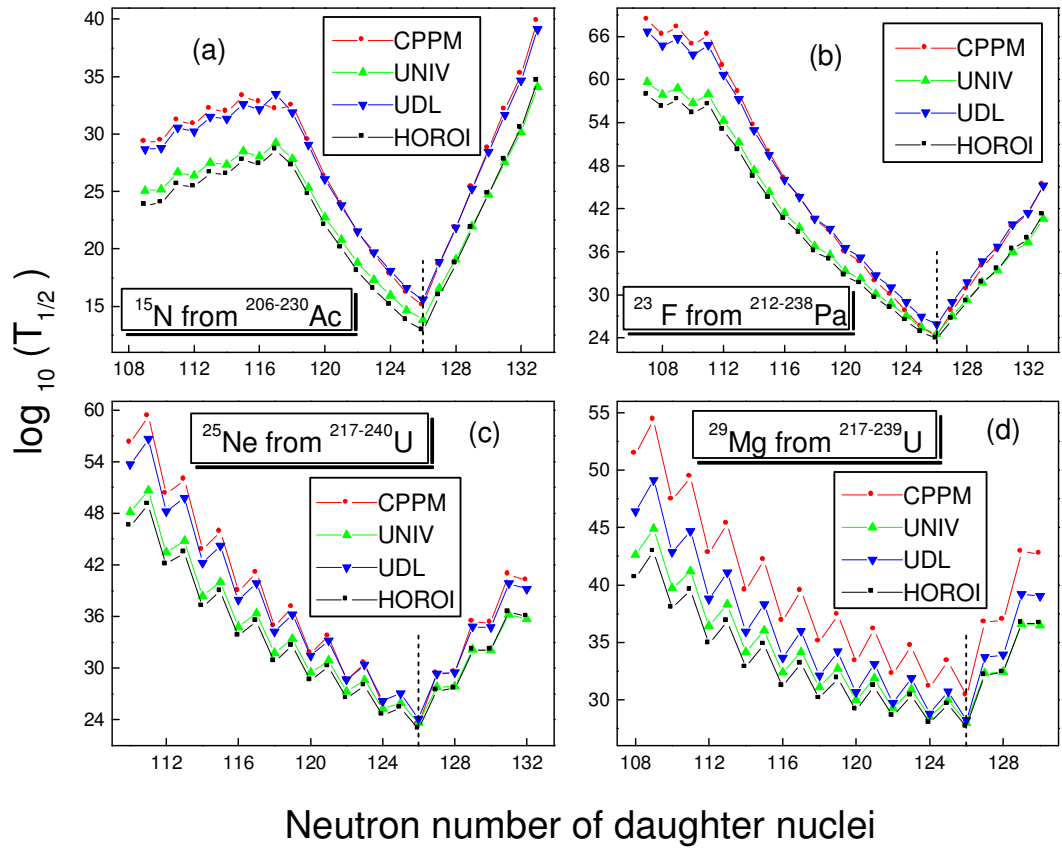

Fig.7. Plot of the computed $\log_{10}(T_{1/2})$ values vs. neutron number of daughter for the emission of clusters $^{15}$N, $^{23}$F, $^{25}$Ne and $^{29}$Mg respectively from Ac, Pa and U isotopes. $T_{1/2}$ is in seconds.

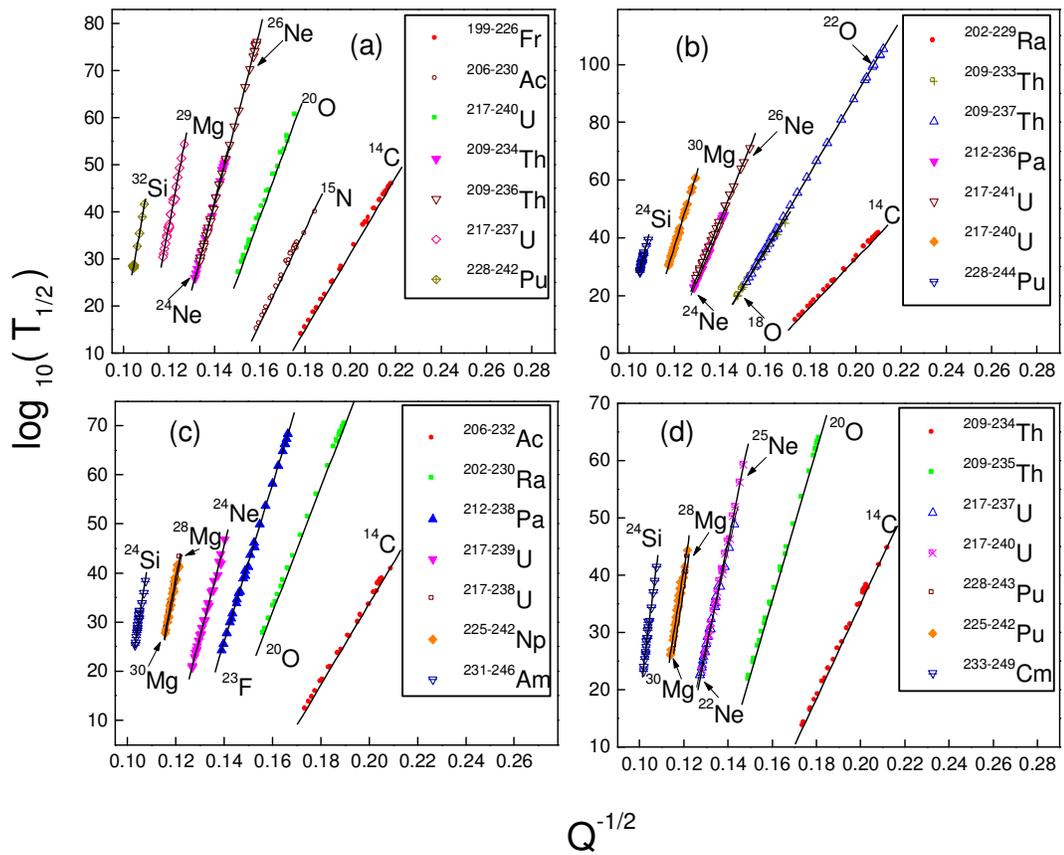

Fig.8. Geiger – Nuttall plot of $\log_{10}(T_{1/2})$ values vs $Q^{-1/2}$ for various clusters from different parents. $T_{1/2}$ is in seconds.